\documentclass[epsfig,a4,oneside,reqno,11pt,psfrag,portrait]{amsproc}
\usepackage{amsfonts,amssymb,amsmath,latexsym}%{amssymb}
\usepackage{enumerate}
\usepackage{a4wide}
\usepackage[english]{babel}

\newtheorem{thm}{\bf Theorem}[section]
\newtheorem{prop}[thm]{\bf Proposition}

\newtheorem{lemma}[thm]{\bf Lemma}

\usepackage[dvipsone]{color}
\usepackage[dvips]{graphicx}

\definecolor{blue2}{rgb}{0,0,0.6}
\definecolor{turck}{rgb}{0,0.5,0.5}
\definecolor{sturck}{rgb}{0,0.9,0.8} % osservare che numeri grandi schiariscono !?
\definecolor{orange}{rgb}{1,0.4,0}
\definecolor{bord}{rgb}{1,0.3,0.3}
\definecolor{violet}{rgb}{0.4,0.2,0.8}
\definecolor{sblue}{rgb}{0.2,0.4,1.}
\definecolor{yellowa}{rgb}{1.,0.8,0}
\definecolor{greena}{rgb}{0.2,0.8,0}
\definecolor{greenb}{rgb}{0,0.5,0.2}
\definecolor{greenc}{rgb}{0.1,0.7,0.3}
\definecolor{greend}{cmyk}{0.7,0,0.3,0}
\definecolor{greene}{cmyk}{0.7,0,0.3,0.1}
\definecolor{greya}{cmyk}{0,0,0,0.1}
\definecolor{greyb}{cmyk}{0,0,0,0.2}
\definecolor{greyc}{cmyk}{0,0,0,0.3}
\definecolor{greyd}{cmyk}{0,0,0,0.4}
\definecolor{greye}{cmyk}{0,0,0,0.5}
\definecolor{greyf}{cmyk}{0,0,0,0.6}
\definecolor{greyg}{cmyk}{0,0,0,0.7}
\definecolor{greyh}{cmyk}{0,0,0,0.8}
\definecolor{greyi}{cmyk}{0,0,0,0.9}
\newcommand{\mintwo}[2]{\min_{\substack{#1 \\ #2}}} % min with 2 lines
\newcommand{\sumtwo}[2]{\sum_{\substack{#1 \\ #2}}} % sum with 2 lines
\newcommand{\al}{\alpha}

\newcommand{\La}{\Lambda}

\newcommand{\ga}{\gamma}
\newcommand{\Ga}{\Gamma}

\newcommand{\si}{\sigma}

\long\def\notes#1{\ifinner
         {\tiny #1}
         \else
          \marginpar{\protect\tiny #1}%
          \fi}%

\newcommand{\dis}{\displaystyle}
\newcommand{\und}{\underline}

\newcommand{\Zz}{\mathbb{Z}}

\newcommand{\cC}{\mathcal{C}}

\newcommand{\cR}{\mathcal{R}}

\newcommand{\Ii}{\text{\bf 1}}
\newcommand{\dist}{\mathrm{dist}} % mathrm distance
\newcommand{\ab}{c}

\newcommand{\uT}{\und{T}}

\newcommand{\vxi}{\vec\xi}
\newcommand{\h}{h}
\begin{document} %##################################################################
%CFMPf.tex -:- 25 novembre 2002 ore 10.08 -:-\vskip 2.5cm \noindent
\title[geometry of contours in $d=1$] %[ {[CFMPf.tex -:- 25 novembre 2002 ore 10.08 -:-]} ]
{Geometry of contours and Peierls estimates\\ in d=1 Ising models
with long range interactions}

\author{M. Cassandro}
\address{Marzio Cassandro,
Dipartimento di Fisica, Universit\`a di Roma La Sapienza and INFM Sezione \newline
\indent di Roma,\newline \indent 00185 Roma, Italy}
\email{Marzio.Cassandro@roma1.infn.it}
%\thanks{Research partially supported by
%MURST and INFM Sezione di Roma}

\author{P.A. Ferrari}
\address{Pablo Augusto Ferrari,
Departamento de Estatistica, Universidade de S\~{a}o Paulo, Brazil}
\email{pablo@ime.usp.br}
%\thanks{{P.A.Ferrari thanks kind hospitality at Università di Roma Tor Vergata where this paper
%was conceived.\newline \indent P.A.Ferrari thanks partial support of (CNR?), FAPESP, CNPq and PRONEX.
%}}%

\author{I. Merola}
\address{Immacolata Merola,
Dipartimento di Matematica, Universit\`a
di Roma Tor Ver\-ga\-ta,\newline \indent 00133 Roma,
Italy}
\email{merola@mat.uniroma2.it}
%\thanks{Research partially supported by
%MURST and NATO Grant PST.CLG.976552}

\author{E. Presutti}
\address{Errico Presutti,
Dipartimento di Matematica, Universit\`a
di Roma Tor Ver\-ga\-ta, \newline \indent00133 Roma,
Italy}
\email{presutti@mat.uniroma2.it}
%\thanks{Research partially supported by
%MURST and NATO Grant PST.CLG.976552}

%risp.: "Phase transition (general)",
%      "Classical equiibrium statistical mechanics(general)",
%        "Lattice systems(Ising, dimmer,...)",
% 2000 MSC --- campiare il file amsproc che scrive 1991 !!!!!!

\begin{abstract}
Following Fr\"ohlich and Spencer, \cite{FS}, we study one
dimensional Ising spin systems with ferromagnetic, long range
interactions  which decay as $|x-y|^{-2+\alpha}$, $0\leq
\alpha\leq 1/2$. We introduce a geometric description of the spin
configurations in terms of triangles which play the role of
contours and for which we establish Peierls bounds. This in
particular yields a direct proof of the well known result by Dyson
about phase transitions at low temperatures.

\hfill\break \phantom{a}\hfill\break \phantom{a}\hfill\break Key words: Ferromagnetic,
long range interactions, Phase transitions, Contours, Peierls estimates

\end{abstract}
\maketitle

\setcounter{equation}{0}
\section{Introdution}

    \vskip .5cm

A rigorous proof of liquid-vapor phase transitions is a long
standing challenge for mathematical physicists. A clear
understanding of the phenomenon goes back to  van der Waals, but a
mathematically consistent theory is still lacking. Lebowitz, Mazel
and Presutti, \cite{LMP}, have tried to capture van der Waals
ideas by considering an Hamiltonian which has a term given by an
 attractive, two body
Kac potential. The effort was to study the model without taking the Kac scaling
parameter $\ga\to 0$, as in the original works of Kac, Uhlenbeck and Hemmer,
\cite{KUH}, and Lebowitz and Penrose, \cite{LP}. Technically, the idea was to study
the system as a perturbation of mean field, which corresponds to the limit case
$\ga=0$, and to adapt to such a context the Pirogov-Sinai theory of finite temperature
perturbations of ground states. To carry through the program, one needs a good control
of an approximate model where the Kac potential term in the hamiltonian is replaced by
a self-consistent, external one body field, whose intensity depends on the true value
of the order parameter [the particles density] at equilibrium.  In the continuum, the
hamiltonian cannot consist of just the attractive
 Kac potential (as in Ising models with Kac potentials) and a
repulsive force is needed to prevent a collapse of matter.  The
natural choice (as proposed originally by Kac et al.) is then to
add a hard core interaction, but, at the required values of the
particles density, the cluster expansion results for the system
with only hard cores are not valid and the implementation of the
Pirogov-Sinai methods collapses. In \cite{LMP} the problem has
been avoided by using repulsive forces which are also given by Kac
potentials, in particular four-body positive interactions. The
escamotage is physically not totally satisfactory, as the phase
transition should arise from a competition between the short range
repulsive and the much longer range attractive inter-molecular
forces. Several efforts to extend \cite{LMP} to such a context and
in particular to the model with hard core plus attractive two body
Kac potentials have failed.

There is however some margin  left if we restrict to one
dimensions, because the pure hard rods system is isomorphic to an
ideal gas. Unfortunately, there is a price to pay: to have a phase
transition in $d=1$,  we need to consider long range forces
(potentials which decay as $|x-y|^{-2+\alpha}$, $\alpha \in
[0,1)$) which are not covered by the traditional Pirogov Sinai
theory. Prior to \cite{LMP}, the problem of phase transition in
the continuum in $d=1$ with such long range interactions  had
already been considered by Johansson, \cite{Johansson,
Johansson2}, who studied the system in the canonical ensemble,
proving phase transition for the thermodynamic potentials. The
existence of distinct DLR measures at the proper values of
chemical potential and temperature remains however open.

The Pirogov-Sinai theory  seems the natural way to answer these
questions, as it provides powerful tools for investigating phase
transitions at low temperatures and at low effective temperatures
as well, with a quite satisfactory description of systems in
dimensions larger or equal to two. In view of the desired
applications to continuum particle models, our mid-term program is
to extend Pirogov-Sinai to one dimensional spin systems with long
range interactions. The content of this paper will be the
definition of contours and the establishment of Peierls estimates,
as a preliminary step in this direction.  After the papers  by
Dyson, \cite{Dyson1}, \cite{Dyson2},  on a model with hierarchical
interactions (which, by ferromagnetic inequalities, prove  phase
transitions in Ising systems as well), we find in the literature
the fundamental paper by Fr\"ohlich and Spencer, \cite{FS}, where
the critical case $\alpha=0$ is studied by deriving Peierls
estimates for suitably defined contours. A further step forward
has then been done by Imbrie, \cite{imbrie}, who proved the
validity of the cluster expansion for this gas of contours.  A
different approach, based on inequalities, has instead been
followed by D\"{u}mcke and Spohn, \cite{spohn1}, and Spohn,
\cite{spohn2}, to prove  phase transitions for systems of $\pm 1$
spins on $\mathbb R$ with long range interactions, $\alpha\in
[0,1)$. The results were used in the analysis of ground states for
some quantum systems.

In this paper  we revisit  Fr\"ohlich and Spencer \cite{FS} and
extend  it to the case $\alpha\in (0,1/2]$.  In particular we
prove that the probability of occurrence of a droplet of the
opposite phase is depressed at least by $c \exp\{-\beta \zeta
L^\al\}$, $c$ and $\zeta$ positive constants, $L$ the length of
the droplet. The analogy with $d>1$ where the bound goes as
 $c \exp\{-\beta \zeta
L^{(d-1)/d}\}$, is evident (our proof applies essentially
unchanged through $\al=0$, where it yields the bound  $c
\exp\{-\beta \zeta \ln L\}$, loosing however the analogy with
$d>1$).   Comforted by these results and the analogy with $d>1$,
we plan, in the future, to extend the analysis to Ising systems
with Kac potentials and then, hopefully, to prove phase
transitions for hard rods with attractive Kac potentials, at least
for $\al>0$.

The bibliography on the subject should also include the papers,
\cite{Aiz-Ch-Ch-Newman,Aiz-Newman,marchetti,Newman1,Newman2},
which refer to $d=1$, long range percolation. In fact, using the
FK representation, the results can be transferred  to Ising
systems, but it is not clear whether the approach could extend to
the continuum particle systems where ferromagnetic inequalities
are absent.

Thus, the model we consider here is an Ising ferromagnet on a one
dimensional lattice, with total energy
\begin{eqnarray}\label{1.1}
  && h(\si)=  \frac{1}{2}\sum_{x,y\in \Zz}J(|x-y|)\text{\bf 1}_{\si(x)\ne\si(y)} \\
 \label{1.2}&& \mbox{} \hskip.3cm J(n)=
  \begin{cases}
    J(1)>> 1& %\text{if } n=1
    \\
    \\
    \frac{\dis{1}}{\; \; \dis{n^{2-\al}}} & \text{if } n>1.
  \end{cases}
    \hskip2.1cm   \mbox{with } \hskip.3cm  \al\in [0,1/2]
\end{eqnarray}
which will be studied at equilibrium with $\beta\gg 1$.  In the
sequel, for notational convenience, we restrict $\alpha\in
(0,1/2]$, the analysis of the case $\alpha=0$ is analogous and
treated in Appendix \ref{app:A}, \ref{app:E} and \ref{app:F}.

 In this paper we will show that the equilibrium configurations for
the system associated to the hamiltonian (1.1) can be described in
terms of contours whose weights satisfy a Peierls bound. These
contours (as in Pirogov-Sinai) are defined as regions which
collect close-by deviations from the ground states. The Peierls
bound follows from the fact that the excess energy of the
associated interfaces is bounded from below proportionally to the
size of the region to a positive power.  To illustrate this point
consider the simple case of three contiguous intervals $B^-$, $A$
and $B^+$. Let $A$ be of size $L$ and $B^{\pm}$ of size larger or
equal to $L$ and call $\mathcal{C}$ the set of configurations s.t.
$\sigma=+1$ for all sites belonging to A and $\sigma=-1$ for all
sites belonging to $B^{\pm}$. An explicit calculation, see
Appendix \ref{app:A}, shows that for all configurations in
$\mathcal{C}$ the variation of energy obtained by flipping the
spins inside $A$ (thus getting all spins equal to $+1$ in $A\sqcup
B^+\sqcup B^-$) is bounded from below by $\zeta_\alpha
L^{\alpha}$, with $\zeta_\alpha>0$ for $\alpha\in (0,1/2]$ and if
$J(1)$ is large enough.
%By introducing a notion of contours it is possible to generalize the argument
%and to extend to $1$-dimensional long range systems the full machinery developed in
%the last decades for Ising spin systems in dimensions larger or equal to $2$.

In Section \ref{sec:2} we give a graphical description of a spin
configuration in terms of a configuration of triangles, which
allows to introduce the notion of internal and external interfaces
(like in $d>1$ dimensions), see Fig. \ref{fig:seconda} in Section
\ref{sec:2}.

In Section \ref{sec:3} we introduce the notion of contours as clusters of nearby
triangles and  prove Peierls bounds for their energy.   Our definition is very similar
to that in \cite{FS}, but our aim is to get  a geometric representation of the
contours   more explicit and better suited for further generalizations.

In Section \ref{sec:4} we prove that for $\beta$ large enough the
Peierls estimates on the energy of contours enable to control
their entropy.

The approach we use can be generalized to a larger class of long range attractive
forces where the assumption $J(1)\gg 1$ is dropped and $\alpha \in (0,1)$.
 We will discuss this point in a
forthcoming paper together with a characterization of the typical
configurations for slow decreasing ferromagnetic Kac potentials.
As mentioned the ultimate goal is the extension to a one
dimensional system of hard-core particles interacting via such
long range attractive forces, but at the moment we have not yet
concrete results in this direction.

\vskip 2.5cm \noindent
\setcounter{equation}{0}

 \section{Spin and triangle configurations}
 \label{sec:2}

We will consider in this paper homogeneous boundary conditions, i.e.\  the spins in
the  boundary conditions are either all $+1$ or all $-1$. By the spin flip symmetry,
we may and will restrict to the former, so that we will only study configurations
$\si= \{\si_x, x\in\mathbb Z\} \in \mathcal X_+$, namely such that $\si_x=1$ for all
$|x|$ large enough. Our aim here is to recover a picture as in $d>1$, where the
configurations are described by a collection of interfaces.  In one dimensions, an
interface at $(x,x+1)$ means $\si_x\si_{x+1}=-1$.  The precise location  of the
interface in the interval is immaterial and we will use it to our advantage by
choosing a point in each interval $(x+1/2)\pm 1 /100$, $x\in \mathbb Z$, with the
property that for any four distinct points $r_i$, $i=1,..,4$, $|r_1-r_2|\neq
|r_3-r_4|$.  We suppose the choice done once for all, so that hereafter an interface
point between $x$ and $x+1$ is uniquely fixed.

Any interface point, by its definition, represents a change of
phase so that after the first interface point (coming from the
left), the second one corresponds to a reestablishing of the
original phase, and so on.  However, this is not the most
convenient way to look at the spin configurations. Our
construction is similar to that in \cite{FS} (where interface
points were called spin flip points) and it is based on suitably
coupling together pairs of interface points. To this end we will
use the criterion of minimal distance, which will be made
geometrically intuitive by using   a graphical representation
where each spin configuration is mapped into a set of triangles.
The endpoints of the triangles will be the pairs of coupled
interface points.

Due to the above choice of the boundary conditions, any $\si \in
\mathcal X_+$ has a finite, even number of interface points.  We
then let each interface point evolve into two trajectories
represented in the $(r,t)$ plane by the two lines $r\pm t, t\ge
0$.  We have thus a bunch of growing v-lines each one emanating
from an interface point.  Once two v-lines meet, they are frozen
and stop their growth, while the others are undisturbed and keep
growing. Our choice of the location of the interface points ensure
that collisions occur one at a time so that the above definition
is unambiguous.
%The left side of Fig. \ref{fig:prima-a} %1a
%shows an example of collisions.

%
%\vskip .5cm \noindent
%\begin{figure}[h]%h=here t=top b=bottom p=Page of floats
%\centering \resizebox{30cm}{!} {\rotatebox{-0}{\includegraphics{coll-ultima-a.eps}}}
%\caption{Examples of v-lines growth (left) and triangles (right). }
%\label{fig:prima-a}
%\end{figure}

The collision of two points is represented graphically in the $(r,t)$ plane by a
triangle whose basis is the line joining the two interface points and whose sides are
the two arms of the v-lines which enter into contact at the time of collision.
%, see the left side of Fig.1.
Triangles will be usually denoted by $T$ and we will write
 \begin{eqnarray}
&& |T | = \text{ cardinality of $T\sqcap \mathbb Z$},\qquad \text{dist$(T,T') =$
cardinality of $I\sqcap \mathbb Z$,}
%\\&&\hskip3cm \text{
%$I$ the interval between $T$ and $T'$}
   \label{n}
 \end{eqnarray}
where $I$ is the interval between $T$ and $T'$ if $T$ and $T'$ are
disjoint;  if $T$ and $T'$ are one contained in the other (no
other possibility may arise in the above construction) then $I$
denotes the minimal interval between the two.

We have thus represented a configuration $\si \in \mathcal X_+$ as
a collection $\und T=(T_1,..,T_n)$ of triangles in the $(r,t)$
plane.  The set of configurations of triangles obtained in this
way are denoted by $\{\und T\}$, and the above construction
defines a one to one map from $\mathcal X_+$ onto $\{\und T\}$. It
is easy to see that a triangle configuration $\und T$ belongs to
$\{\und T\}$ iff for any pair $T$ and $T'$ in $\und T$
 \begin{equation}
   \label{2.1}
\text{dist}(T,T') \ge \min\big\{|T|,|T'|\big\}
 \end{equation}

\begin{figure}[h]%h=here t=top b=bottom p=Page of floats
\centering \resizebox{7cm}{!} {\rotatebox{-90}{\includegraphics{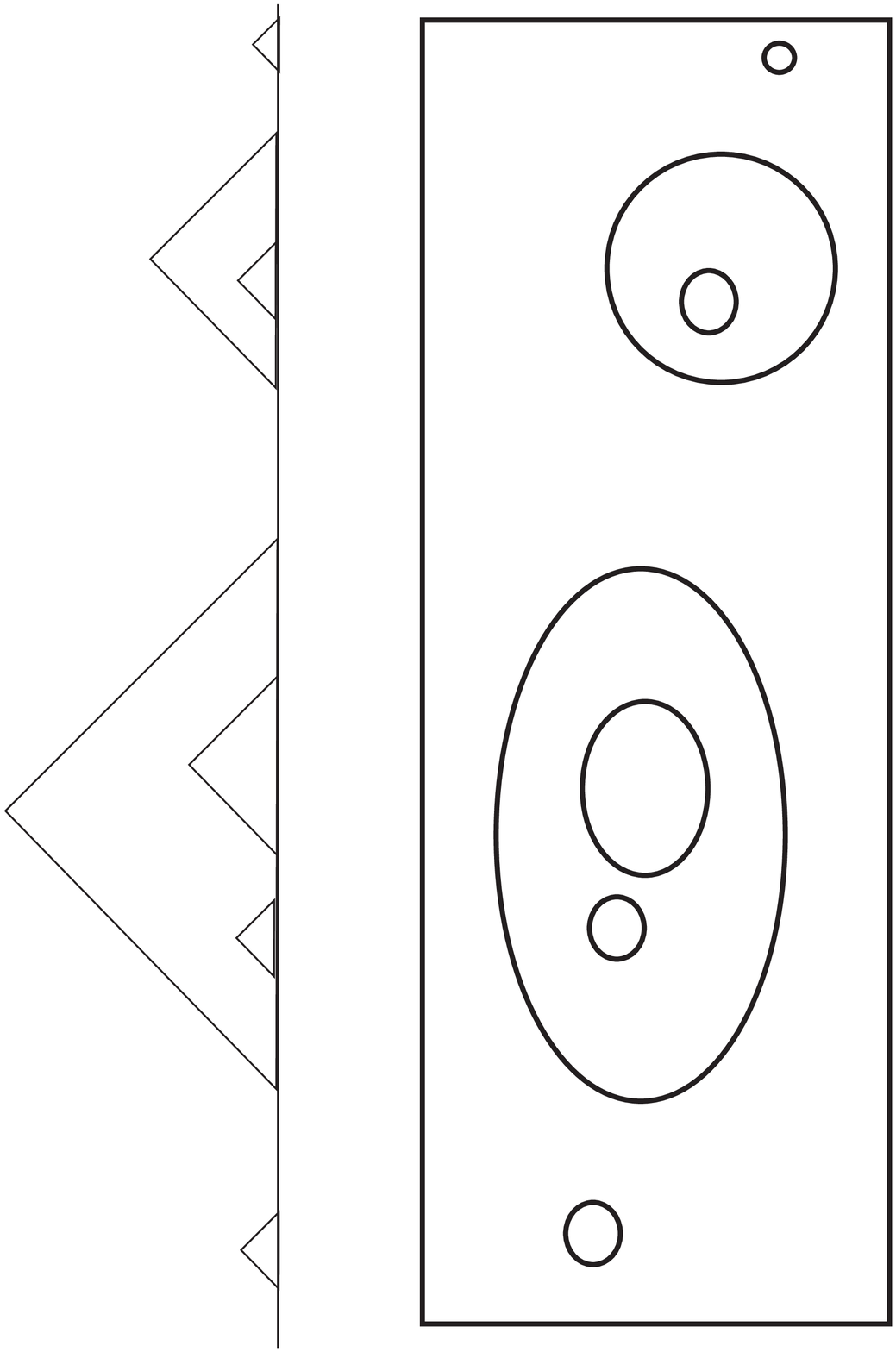}}}
\caption{} \label{fig:seconda}
\end{figure}

The two endpoints of a triangle play the role which has the interface in higher
dimensions and we thus have, also in $d=1$, a notion of external and internal
interfaces. (see Fig. \ref{fig:seconda})

The above construction is taken from a model for $d=1$ coarsening,
see Derrida, \cite{Derrida},   Carr and Pego, \cite{Pego} and also
some old, unpublished notes of two of us (P.F and E.P). Coarsening
in $d=1$ is extremely slow  and it is often a good approximation
to say that in a given sequence of intervals of alternating
phases, the shortest one disappears first, while all the others
are unchanged.  The dynamics is then described in terms of
triangles by calling the first interval which disappears as the
basis of the smallest triangle and then iterating the procedure.
The interesting setup when studying coarsening is to have
initially infinitely many phase changes and one of the aims is to
understand if and which are the self similar structures which
emerge from the triangles picture. Here our task is simpler, we
have finitely many phase changes and want to prove that each one
of them has a small Gibbs weight.

Writing
 \begin{equation}
   \label{2.2}
H(\und T)= h(\si),  \qquad \si \in \mathcal X_+
\;\Leftrightarrow\; \und T \in \{\und T\}
 \end{equation}
and calling $\und T=(T_1,..,T_n)$ with $|T_i|\le |T_{i+1}|$, we have
 \begin{equation}
   \label{2.3}
H(\und T)= H\big(T_1 \,\big|\,\und T\setminus T_1\big)+ H\big(\und T\setminus
T_1\big),\qquad H\big(\und S\,\big|\,\und T\big):= H(\und S\sqcup\und T)-H(\und T)
 \end{equation}
In fact if $\und T\in \{\und T\}$, and $T\in \und T$, then $\und
T\setminus T$  obviously satisfies \eqref{2.1} and therefore it is
in $\{\und T\}$.  $\und T\setminus T$ is obtained from the
configuration $\si$ corresponding to $\und T$ by flipping all the
spins inside the basis of $T$. By iteration,
 \begin{equation}
   \label{2.4}
H(\und T)= \sum_{i=1}^n H\big(T_i \,\big|\,\und T\setminus [T_1
\sqcup\cdots \sqcup T_{i}]\big)
 \end{equation}

 \vskip.5cm

 \begin{lemma}
 \label{lemma2.1}
 For any $i$,
 \begin{equation}
   \label{2.5}
H\big(T_i \,\big|\,\und T\setminus [T_1 \sqcup\cdots \sqcup
T_{i}]\big) \ge W(|T_i|)
 \end{equation}
where
% \begin{equation}
%   \label{2.6}
%W(|T_i|):= \sum_{x\in T_i\sqcap \mathbb Z}\Big(\sum_{y\in
%I_i^{\pm}} J(|x-y|) -\sum_{y\in I_i'}J(|x-y|)\Big),\qquad I_i'=
%\mathbb Z \setminus (T_i\sqcup I_i^+\sqcup I_i^-)
% \end{equation}
%
 \begin{equation}
   \label{2.6}
W(L)= \sum_{x=1}^{L}\left(\sumtwo{y\in [L+1,2L]}{y\in
[-L+1,0]}J(|x-y|)- \sumtwo{y\in [2L+1,\infty]}{y\in
[-\infty,-L]}J(|x-y|)\right)
 \end{equation}
%
%
%
%
%and $I_i^{\pm}$ are the two intervals  in $\mathbb Z$ to the right
%and left of $T_i$, consisting each of $|T_i|$ sites.

 \end{lemma}

 \vskip.5cm

{\bf Proof.}  Call  $I_i^{\pm}$ the two intervals  in $\mathbb Z$
which are to the right and to the left of $T_i$, each one
consisting of $|T_i|$ sites.
 There is no interface point inside $T_i$, and inside
$I_i^{\pm}$ as well, because $|T_i|$ is the minimal length in
$\und T\setminus [T_1 \sqcup\cdots \sqcup T_{i-1}]$ and all $T_j$,
$j>i$, have distance from $T_i$ which is $\geq |T_i|$. Then, if
$\si$ corresponds to $\und T\setminus [T_1 \sqcup\cdots \sqcup
T_{i-1}]$, the spins in $T_i\sqcap \mathbb Z$ are all equal to
each other and  opposite to those in $I_i^{\pm}$. Instead, in the
configuration $\si'$ which corresponds to $\und T\setminus [T_1
\sqcup\cdots \sqcup T_{i}]$, the spins are all the same in
$(T_i\sqcap \mathbb Z)\sqcup I_i^+\sqcup I_i^-$. By \eqref{2.3},
$H\big(T_i \,\big|\,\und T\setminus [T_1 \sqcup\cdots \sqcup
T_{i}]\big) = h(\si)-h(\si')$, so that \eqref{2.5} and the lemma
are proved.

\qed

% intersect, the configuration
%$\si$ corresponding to $\und T$ has all spins equal to each other
%inside $T_1$. Calling $I_1^{\pm}$ the two intervals to the right
%and left of $T_1$ with same number of points as in $T_1\sqcap
%\mathbb Z$, by \eqref{2.1} and since $|T_1|$ is the smallest
%length, also the spins in each one of the two intervals
%$I_1^{\pm}$ have constant sign.  The two interface points of $T_1$
%allow to conclude that the spins in the whole $I_1^{+} \sqcup
%I_1^{-}$ have constant sign which is opposite to that of the spins
%in $T_1$.
% The configuration $\si'$ corresponding to $\und T\setminus T_1$ is
%obtained by flipping the spins inside $T_1$ which therefore remain equal to each other
%and to those in $I_1^{\pm}$.
%
%
%
%
%
%We are going to prove that f
%
%
%{\bf Proof of \eqref{2.5}.}  By the same argument as below \eqref{2.3}, in the
%configuration $\si$ corresponding to $\und T\setminus [T_1 \sqcup\cdots \sqcup
%T_{i-1}]$ the spins in $T_i\sqcap \mathbb Z$ are equal to each other and have opposite
%sign than those in  $I_i^{\pm}$, while in the configuration $\si'$ corresponding to
%$\und T\setminus [T_1 \sqcup\cdots \sqcup T_{i}]$ all spins are the same in $T_i\sqcup
%\mathbb Z$, $I_i^+$, $I_i^-$. Since $H\big(T_i \,\big|\,\und T\setminus [T_1
%\sqcup\cdots \sqcup T_{i}]\big) = h(\si)-h(\si')$, \eqref{2.5}-\eqref{2.6}  follow
%from the claim.

\vskip1cm

In Lemma \ref{lem:primo}  it is proved that for $J(1)$ large
enough, there is $\zeta>0$ so that
 \begin{equation}
   \label{2.7}
W(L)\ge \zeta \h_\al(L) \hskip1cm \text{where} \hskip1cm
 \h_\al(L):= \begin{cases}
     L^\alpha & \al\in (0,1/2] \\
    \ln L + 4  & \al=0.
  \end{cases}
 \end{equation}
 in the sequel we fix our attention on the case $\al\in (0,1/2]$, and discuss the case
$\al=0$ in Appendix %\ref{app:F}.
\ref{app:A}, \ref{app:E} and \ref{app:F}.
%\begin{equation}
%   \label{2.7}
%W(L)\ge \zeta L^\alpha
% \end{equation}
Thus
 \begin{equation}
   \label{2.8}
H(\und T)\ge \zeta \sum_{i=1}^n |T_i |^\alpha,\qquad  \und T=
(T_1,..,T_n)
 \end{equation}
The inequality must be seen as an analogue of the Peierls estimate in $d>1$ where the
excess energy of a configuration of interfaces is bounded from below proportionally to
the surface area of such interfaces. Since $|T_i|$ is the volume surrounded by the
interface,   $\alpha$ is identified to the ratio $(d-1)/d$, with $d$ an ``effective
dimension'' of the system.

This is however only an analogy. To really implement a Peierls
bound in our setup, we need to ``localize the estimates'', being
able to compute the weight of a given triangle in a generic
configuration. The previous bound was easy, because we could
estimate successively the weights of the triangles in the same
order as their lengths. If we want to bound the energy of a
generic triangle $T$ in  configuration $\und T$, $|T|$ may not be
the smallest length so that we are confronted with cases where
there are other triangles in $T\sqcup I^+\sqcup I^-$ (see Lemma
\ref{lemma2.1} for notation). Indeed, we could add to $T$ smaller
triangles $T'$ in $T\sqcup I^+\sqcup I^-$ without violating
\eqref{2.1}. Our approach will be $\bullet$\; to ``connect''
triangles if they are ``dangerously close'' to each other,
$\bullet$\; to define contours as ``connected clusters'' of
triangles and  $\bullet$\; to compute probabilities of contours
rather than of single triangles. %The implementation of such an
%idea will involve a hierarchical structure to define the
%connections among triangles in the cluster.
 To compute the
probability of a contour, we first order increasingly the
triangles in the contour, according to their lengths. Then the
previous argument can be generalized, exploiting the fact that the
triangles which are not in the contour %and thus belong  to
%distinct contours,
are ``sufficiently far away'' (by the way contours are defined).
In the next section we will see how triangles can be clustered
into contours and then extend Lemma \ref{lemma2.1} to contours,
thus concluding the analysis of the energy of contours; in Section
\ref{sec:4} we will prove entropy bounds (on the number of
contours), which show that for $\beta$ large enough, energy wins
against entropy.

\vskip3cm

\setcounter{equation}{0}

 \section{Contours and Peierls estimates}
 \label{sec:3}

In Subsection \ref{subsec:4.1} we will define a function $\mathcal
R$ which associates to any configuration $\und T\in\{\und T\}$ a
configuration $\{\Ga_j\}$ of contours, each $\Ga_j$ being a subset
of  triangles in $\und T$.  The crucial point in the definition is
that the triangles in a contour are ``close to each other'', while
all the other triangles are ``far away''; using such a property we
will be able to extend to contours the energy estimate of the
previous section, thus deriving the  Peierls estimates of
Subsection \ref{subsec:3.2}.  In Subsection \ref{subsec:3.3} we
will recall the classical argument for existence of a phase
transition, using the Peierls bound proved in Subsection
\ref{subsec:3.2} and the entropy estimates which will be proved in
Section \ref{sec:4}.

 \subsection{Contours}
 \label{subsec:4.1}

A contour $\Ga$ is a  collection $\und T $ of triangles ($\und T$
in this Section will always, and sometimes tacitly, denote an
element in $\{\und T \}$) joined together by a hierarchical
network of connections, under which all the triangles of a contour
become mutually connected. The structure has a self similar
property which we will exploit when counting the contours. The
coarsest picture of a contour $\Ga$ is the pair $\{T(\Ga),
|\Ga|\}$, $T(\Ga)$ a triangle, $|\Ga|$
%(also denoted by
%$|T(\Ga)|$)
its mass. $T(\Ga)$ is the triangle whose basis is the
smallest interval which contains all the triangles of the contour,
the right and left endpoints of   $T(\Ga)\sqcap \mathbb Z$  are
denoted by $x_{\pm}(\Ga)$.%, $\ell(\Ga) = x_{+}(\Ga)-x_{-}(\Ga)$ is
%the length of its basis.
$|\Ga|$, the mass of the contour, is the
sum of the masses of all the triangles in $\Ga$, the mass $|T_i| $
of a triangle being defined in \eqref{n}.
%Notice that
%$\ell(T(\Ga))$ may be different from $|T(\Ga)|=|\Ga|$.

Our aim is to define an algorithm $\mathcal R(\und T)$ on $\{\und T\}$, which
associates to any configuration $\und T$ a configuration $\{\Ga_j\}$ of contours with
the following properties.

 \vskip1cm

 {\bf P.0} {\it  Let $\mathcal R(\und T)=(\Ga_1,..,\Ga_n)$, $\Ga_i =\{ T_{j,i}, 1\le j
 \le k_i\}$, then $\und T=\{T_{j,i}, 1\le i \le n,\;1\le j
 \le k_i\}$}.

 \vskip.5cm

{\bf P.1} {\it Contours are well separated from each other.} Any pair $\Ga\ne\Ga'$ in
$\mathcal R(\und T)$  verifies one of the following two alternatives. (i):
$T(\Ga)\sqcap T(\Ga')=\emptyset$, in which case
 \begin{equation}
   \label{4.1}
\text{dist}(\Ga,\Ga') > c \min\big\{|\Ga|^3,|\Ga'|^3\big\}
 \end{equation}

%where  $c$  is  as in \eqref{3.9} below and  dist$(\cdot,\cdot)$
%(which in the present case is equal to the distance between
%$T(\Ga)$ and $T(\Ga')$) means distance between the set of
%endpoints of all the triangles in $\Ga$ from the corresponding set
%in $\Ga'$.
{where  $c$  is  as in \eqref{3.9} below and  dist$(\cdot,\cdot)$
 means distance as defined in \eqref{n} between the set of
 all the triangles in $\Ga$ from the corresponding set
in $\Ga'$: \[\dist(\Ga,\Ga'):= \mintwo{T\in\Ga}{T'\in\Ga'}\dist(T,T')\]
(which in the present case is equal to the distance between the two triangles
$T(\Ga)$ and $T(\Ga')$)}

\vskip .5cm \noindent

(ii): $T(\Ga)\sqcap T(\Ga')\ne \emptyset$, then either $T(\Ga)\sqsubset T(\Ga')$ or
 $T(\Ga')\sqsubset T(\Ga)$; moreover, supposing for instance that the former case is verified,
 (in which case we call $\Ga$ an inner contour)
then  for any triangle $T'_i\in \Ga'$, either $T(\Ga)\sqsubset T'_i$ or $T(\Ga)\sqcap
T'_i=\emptyset$; and
\begin{equation}
   \label{4.2}
\text{dist}(\Ga,\Ga') > c |\Ga|^3,\quad \text{if ~$T(\Ga) \sqsubset T(\Ga')$}
%\text{dist}(\Ga,\Ga') \geq c |\Ga|^3,\quad \text{if $T(\Ga) \sqsubset
%T(\Ga')$}
 \end{equation}

 \vskip.5cm

%$\bullet$\; {\it  Independence.} Let $\{T_i\}$ be a configuration
%of triangles all in $\mathbb Z_+$ and $\{\Ga_i\}$ the
%corresponding contours; suppose that each $\Ga_i$ has distance
%from the origin $>c|\Ga_i|^3$.  Then if $\{T'_j\}$ is any
%configuration of triangles all in $\mathbb Z_-$ and $\{\Ga'_j\}$
%the corresponding contours, the contours of the configuration
%$\{T_i,T'_j\}$ are $\{\Ga_i,\Ga'_j\}$.  We will use a stronger
%version of this property in the case of intervals: let $(x,y)$ be
%a finite interval in $\mathbb Z$, $\{T_i\}$ a configuration in
%$(x,y)$ and $\{T'_j\}$ a configuration of triangles with basis
%either in the complement of  $(x,y)$ or containing  $(x,y)$. Let
%$\Ga_i$ and $\{\Ga'_j\}$ be the contours  corresponding to
%$\{T_i\}$ and $\{T'_j\}$. Then if the distance of each $\Ga_i$
%from $x$ and $y$ is $>c|\Ga_i|^3$, the contours of the
%configuration $\{T_i,T'_j\}$ are $\{\Ga_i,\Ga'_j\}$.

 {\bf P.2} {\it  Independence.} Let $\{\und T^{(1)},..,\und
T^{(k)}\}$, be $k> 1$ configurations of triangles; $\mathcal
R(\und T^{(i)})=\{\Ga^{(i)}_j, j=1,..,n_i\}$ the contours of the
configuration $\und T^{(i)}$.  Then, if any distinct pair
$\Ga^{(i)}_j$ and $\Ga^{(i')}_{j'}$ satisfies P.1,
 \begin{equation}
   \label{4.2e}
\mathcal R\big(\und T^{(1)},..,\und T^{(k)}\big)=\{\Ga^{(i)}_j,
j=1,..,n_i; i=1,..,k\}
 \end{equation}

%
% \vskip.5cm
%
%$\bullet$\; {\bf Property 3.} {\it Monotonicity.} Let $\Ga$ be a contour in the
%configuration  $\{T_i\}$ and let  $\{T'_j\}$ be a con\-fi\-gu\-ra\-tion obtained from
%$\{T_i\}$ by the addition of some new triangles. Then  $\{T'_j\}$ has a contour $\Ga'$
%which contains all the triangles of $\Ga$.

 \vskip1cm

It is a nice fact of life that not only P.0, P.1 and P.2 can be
actually  implemented by some algorithm $\mathcal R$, but also
that such an algorithm is unique. In Appendix \ref{app:thm4} we
will prove the following theorem:

 \vskip.5cm

 \begin{thm} [Existence and uniqueness]
 \label{thm4.1}
There is a unique algorithm $\mathcal R(\und T)$ which satisfies P.0, P.1 and P.2.

 \end{thm}

 \vskip1cm

 \subsection{Peierls estimates}
 \label{subsec:3.2}
The idea behind the proof of the Peierls estimates, Theorem \ref{thm3.1} below, is
that the property P.1 will ensure  that the triangles which do not belong to a contour
are so far away that, to leading order, they can be neglected and the bond \eqref{2.5}
can be extended to contours.

 \vskip.5cm

 \begin{thm}
 \label{thm3.1}
Let the constant $c$ in the definition of the contours (see P.1) be so large that
\eqref{3.9} below holds. For any $\und T \in \{\und T\}$, let $\Ga_0\in \mathcal
R(\und T)$, $\und T^{(0)}$ the triangles in $\Ga_0$, $\zeta>0$  as in \eqref{2.7}.
Then
 \begin{equation}
   \label{3.1}
H\big(\und T^{(0)} \,\big|\,\und T\setminus \und T^{(0)}\big) \ge
\frac \zeta 2 \sum_{T\in \und T^{(0)}} |T|^\alpha
 \end{equation}
(for $\al=0$, \eqref{3.1} holds with $|T|^\alpha$ replaced by $\log |T|+4$).

 \end{thm}

 \vskip.5cm
 {\bf Proof.}
 Calling $\und T^{(0)}=(T_1,..,T_k)$, $|T_i|\le |T_{i+1}|$, $i=1,..,k-1$,
 \begin{equation}
   \label{3.2}
H\big(\und T^{(0)} \,\big|\,\und T\setminus \und T^{(0)}\big)= \sum_{i=1}^k H\big(T_i
\,\big|\,\und T\setminus \{T_1 ,\dots, T_{i}\}\big)
 \end{equation}
As a difference with Section \ref{sec:2}, here we may have triangles in $I_i^{\pm}$,
but, by the argument after \eqref{2.3},
 \begin{equation}
   \label{3.2a}
(I_i^+\sqcup I_i^-)\sqcap T_j = \emptyset, \qquad \text{for all $j>i$
$T_j\not\sqsupset T_i$}
 \end{equation}
We also have, calling $\{\Ga_j, j\ge 1\}$,   the other contours of $\und T$, different
from $\Ga_0$,
 \begin{equation}
   \label{3.2b}
(I_i^+\sqcup I_i^-)\sqcap T = \emptyset, \qquad \text{for all $T\in \Ga_j$,
$T\not\sqsupset T_i$,  $j\ge 1$  such that
 $|\Ga_j|\ge |\Ga_0|$}
 \end{equation}
because, by property P.1 of Section \ref{sec:4}, dist$(T_i,\Ga_j) \ge$
dist$(\Ga_0,\Ga_j) \ge c|\Ga_0|^3 \ge |T_i|$.

Finally, using again P.1,
 \begin{equation}
   \label{3.2c}
\text{dist}(T_i, T) > c |\Ga_j|^3, \qquad \text{for all $T\in \Ga_j$, $j\ge 1$ and
such that
 $|\Ga_j|< |\Ga_0|$}
 \end{equation}

With the notation introduced after \eqref{2.5}, and with
 \begin{equation}
   \label{3.4}
A(T_i;\Ga_j)=\;\;\bigsqcup_{T\in \Ga_j,T_i\not\sqsubset T}\;\; T\sqcap \mathbb
Z;\qquad |\Ga| = \sum_{T\in \Ga} |T|
 \end{equation}
we claim that
 \begin{eqnarray}
&& H\big(T_i \,\big|\,\und T\setminus [T_1 \sqcup\cdots \sqcup
T_{i}]\big) \ge W(|T_i|) \nonumber \\&& \hskip1cm - 2 \sum_{M}
\sum_{j=1}^n \text{\bf 1}_{|\Ga_j|=M} \sum_{x\in T_i\sqcap \mathbb
Z} \sum_{y\in I_i^{\pm}} J(|x-y|) \Big(\text{\bf 1}_{y\in
A(T_i;\Ga_j)} + \text{\bf 1}_{x\in A(T_i;\Ga_j) }\Big)
   \label{3.3}
 \end{eqnarray}
To prove \eqref{3.3}, we observe that  the  contribution of $\si_x$ and $\si_y$, ($x$
and $y$ as in \eqref{3.3})  is the same as in $ W(|T_i|)$ whenever $\si_x\si_y=-1$; on
the other hand, if $\si_x=\si_y$  then there must exist a triangle distinct from $T_i$
which contains one site and not the other one.  We thus automatically exclude the
triangles which contain $T_i$, as, by \eqref{2.1}, they will also contain $I_i^{\pm}$;
by \eqref{3.2a}, $(T_{i+1},..,T_k)$ are also excluded. Then \eqref{3.3} follows after
noticing that if $\si_x=\si_y$, the pair $\si_x,\si_y$ contributes with the opposite
sign to the energy as for $W(|T_i|)$, hence the factor 2 in the second term on the
r.h.s.\ of \eqref{3.3}. In  \eqref{3.3} we have also split the sum over all contours
putting together contours with same mass, the mass of a contour $\Ga$ being defined in
\eqref{3.4}.

Call $y_0$ the rightmost point of $\mathbb Z$ in $T_i$, $y_1\in I_i^+$ the point, if
it exists, separated from $y_0$ by $[cM^3]$ sites, $[\cdot]$ the integer part of
$\cdot$. By \eqref{3.2b}-\eqref{3.2c} the following holds: any $\Ga_j$ with
$|\Ga_j|=M$ is such that all its triangles which do not contain  $T_i$ and are to its
right, have their left endpoint to the right of $y_1$.    After changing labels, let
$\Ga_1$ be the contour of mass $M$ with the closest triangle to $y_1$ (and to its
right). The triangles in $\Ga_j$, $j>1$, with mass $M$, cannot be closer than $y_2\in
I_i^+$, where $y_2$ (if it exists) is separated from $y_1$ by $[cM^3]$ sites. By
iteration we define $y_j$, $j>2$, and have that the $n$-th closest contour to $T_i$ of
mass $M$ and to its right, is to the right of $y_j$. Calling $y_n$ the last of such
points in $I_i^+$, we have, for any $x\in T_i$,
 \begin{eqnarray}
&&  \sum_{j=1}^n \text{\bf 1}_{|\Ga_j|=M}  \sum_{y \in I_i^{+}} J(|x-y|)\text{\bf
1}_{y\in A(T;\Ga_j)} \le M\sum_{k=1}^n J(|x-y_k|)
%
%
%
%\nonumber \\&& \hskip1cm - 2 \sum_{M} \sum_{j=1}^n \text{\bf
%1}_{|\Ga_j|=M}  \sum_{x\in T_i; y \in T_i^{\pm}}  J(|x-y|)
%\Big(\text{\bf 1}_{y\in A(T;\Ga_j)} + \text{\bf 1}_{x\in
%A(T;\Ga_j) }\Big)
   \label{3.5}
 \end{eqnarray}
because $J(|x-y|)=J(y-x)$ is a decreasing function of $y$ and the total number of
sites in the triangles of a contour $\Ga$ is not larger than $|\Ga|$ (not necessarily
equal because a triangle might be contained in another one).  Moreover, by
monotonicity,
 \begin{eqnarray}
&&  J(|x-y_k|) \le \frac{1}{[cM^3]} \sum_{y\in (y_{k-1},y_k)} J(|x-y|)
%
%
%
%\nonumber \\&& \hskip1cm - 2 \sum_{M} \sum_{j=1}^n \text{\bf
%1}_{|\Ga_j|=M}  \sum_{x\in T_i; y \in T_i^{\pm}}  J(|x-y|)
%\Big(\text{\bf 1}_{y\in A(T;\Ga_j)} + \text{\bf 1}_{x\in
%A(T;\Ga_j) }\Big)
   \label{3.6}
 \end{eqnarray}
so that
 \begin{eqnarray}
&&  \sum_{j=1}^n \text{\bf 1}_{|\Ga_j|=M}  \sum_{y \in I_i^{+}} J(|x-y|)\text{\bf
1}_{y\in A(T;\Ga_j)} \le \frac{M}{[cM^3]} \sum_{y \in I_i^+ } J(|x-y|)
   \label{3.7}
 \end{eqnarray}
The sum is the same as in  $W(|T_i|)$.  Repeating the same procedure for $T_i$ and
$I_i^-$ we finally get
 \begin{eqnarray}
&& H\big(T_i \,\big|\,\und T\setminus [T_1 \sqcup\cdots \sqcup
T_{i}]\big) \ge W(|T_i|) \Big(1-  \sum_{M} \frac{4M}{[cM^3]}\Big)
   \label{3.8}
 \end{eqnarray}
By choosing $c$ so large that
 \begin{eqnarray}
&&  \sum_{M} \frac{4M}{[cM^3]}\le \frac 12
   \label{3.9}
 \end{eqnarray}
and recalling \eqref{2.7} we then prove the theorem.  \qed

 \vskip1cm

 \subsection{Phase transitions}
 \label{subsec:3.3}

To prove phase transitions we follow the well known argument for $d>1$. Let $\La$ be
an interval containing the origin, $\mu_{\La}^+$ the Gibbs measure in $\La$ with $+$
boundary conditions.  Then
 \begin{equation}
   \label{3.10}
\mu_{\La}^+(\si_0=-1) \le \mu_{\La}^+\Big( \{0\in \Ga\}\Big)
 \end{equation}
where $\{0\in \Ga\}$ denotes the event that there is a contour $\Ga$ which has a
triangle $T$  which contains the origin.  Then
 \begin{equation*}
%   \label{3.10}
\mu_{\La}^+\Big( \{0\in \Ga\}\Big) = \frac{1}{Z_{\La}^+} \sum_{\Ga \ni 0}
\;\;\sum_{\und T: \Ga\in \mathcal R(\und T)} e^{-\beta H(\und T)}
 \end{equation*}
Calling $\und T^{(0)}$ the collection of triangles in $\Ga$, $\mathcal R(\und
T^{(0)})=\Ga$,
 by Theorem \ref{thm3.1},
 \begin{equation}
   \label{3.13}
 e^{-\beta H(\und T)} \leq e^{-\beta H\big(\und T\setminus \und T^{(0)}\big)}\;
 w_{\zeta \beta /2}(\Ga)
 \end{equation}
where, for $b>0$,
 \begin{equation}
   \label{4.4}
   w_b(\Ga) := \prod_{T\in \Ga} e^{ - b | T  |^\al}
 \end{equation}
$ w_b(\Ga)$ is called the b-weight of the contour $\Ga$.   Then, using \eqref{3.13},
 \begin{equation}
   \label{3.11}
\mu_{\La}^+\Big( \{0\in  \Ga\}\Big) \le   \sum_{\Ga\ni 0} w_{\zeta \beta /2}(\Ga) =
 \sum_m\;\; \sum_{\Ga: |\Ga|=m, 0\in \Ga}\;
w_{\zeta \beta /2}(\Ga)
 \end{equation}
and, by \eqref{4.5} below, valid for   $\beta$ large enough,
 \begin{equation}
   \label{3.12}
\mu_{\La}^+\Big( \{0\in  \Ga\}\Big) \le 2\sum _m m e^{- \zeta \beta m^\alpha/2}
 \end{equation}
Since the sum starts from $m\ge 1$,  the r.h.s.\ is $<1/2$ if $\beta$ is large enough,
hence the spin flip symmetry is broken  and there is a phase transition.

\vskip3cm

\setcounter{equation}{0}

 \section{Entropy of contours}
 \label{sec:4}

The main result in this section is Theorem \ref{thm4.2} below,
where we prove  \eqref{3.12},  and hence that, for $\beta$ large,
entropy is controlled by energy and a phase transition occurs.

 \vskip1cm

   \begin{thm}
    \label{thm4.2}

For any $b$ large enough and any $m>0$
 \begin{equation}
   \label{4.5}
  \sum_{\Ga: |\Ga|=m, 0\in  \Ga} w_b(\Ga) \le 2 m e^{-b m^\al}
 \end{equation}
   \end{thm}
where $w_b(\Ga)$ has been defined in \eqref{4.4}.

 \vskip.5cm

The theorem is proved in Subsection \ref{subsec:4.6}, by exploiting a self-similarity
property of the contours which is the argument of the next two subsections.

\vskip2cm

 \subsection{An auxiliary branching process}
 \label{subsec:4.3}

Contours  can be described in terms of trees with a self-similar,
hierarchical structure.  We will first describe abstractly the
trees and then relate them to the contours.

The nodes of the tree are ``individuals'' of two species: heavy
triangles, h-triangles in short, and spheres; the h-triangles can
be either black or white.  Only black triangles can procreate and
their offsprings contain at least two h-triangles.  The offsprings
in a branching are ordered, the h-triangles are drawn
sequentially, the spheres, also drawn sequentially, can lie in
each one of the intervals in between two consecutive h-triangles,
but also ``inside'' the white triangles, the latter will be called
``attached'' to the white triangle in  which they are contained.

\begin{figure}[h]%h=here t=top b=bottom p=Page of floats
\centering \resizebox{7cm}{!} {\rotatebox{-0}{\includegraphics{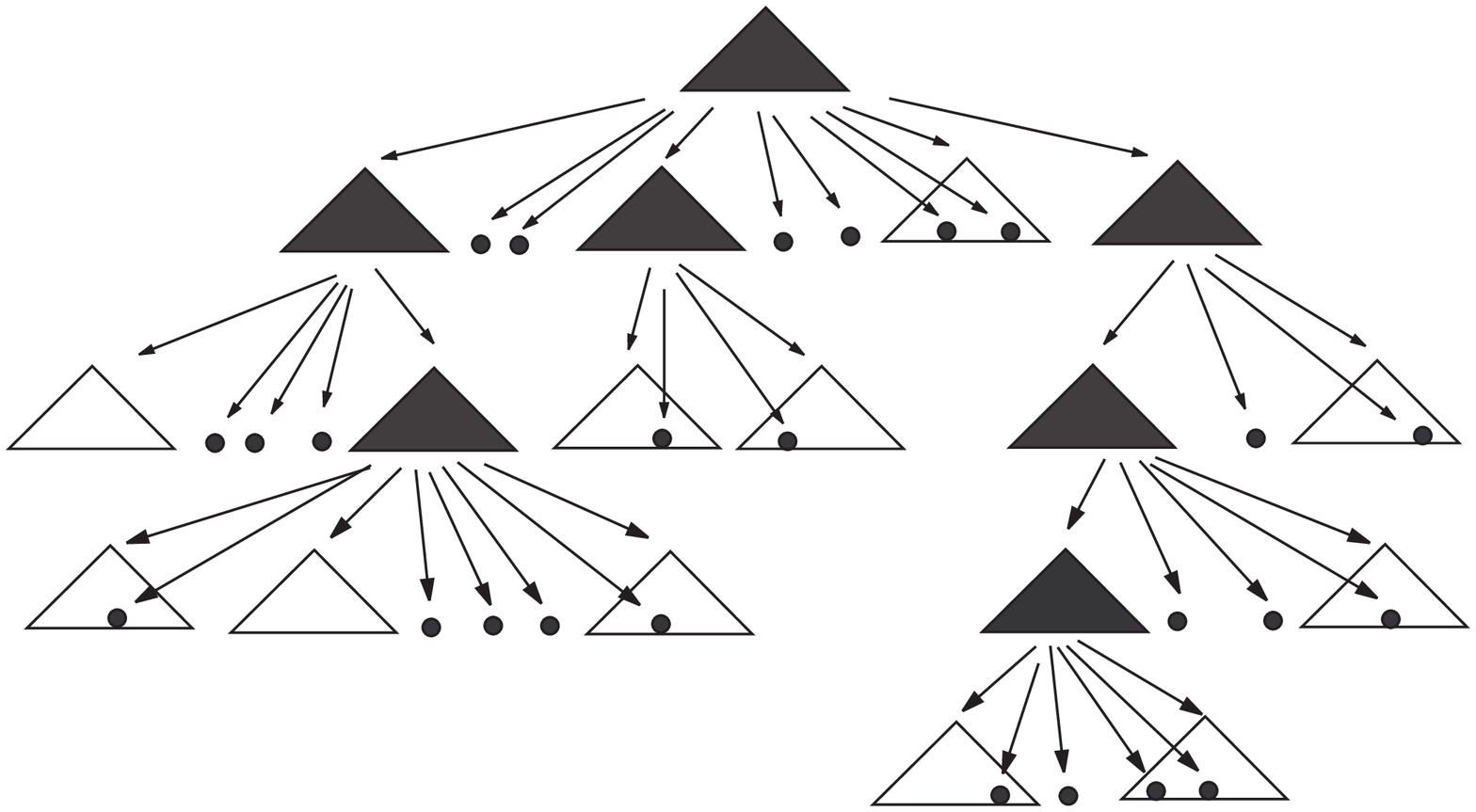}}} \caption{}
\label{fig:tree}
\end{figure}

Finally the tree has a root which consists either of a single
black triangle or  of a single white triangle with possibly
spheres inside the white triangle.  In the second alternative the
tree consists of only its root, as white triangles and spheres
cannot procreate.
 An example of tree is
drawn in Figure \ref{fig:tree}.

We will construct an algorithm which associates to any contour a
tree with the above properties and later use such a correspondence
to prove \eqref{4.5}.  We will in fact organize the sum over
contours in \eqref{4.5} by summing over trees after having summed
over all contours which produce the same tree. The identification
of the nodes of the tree in terms of contours will allow for an
inductive procedure which greatly reduces the complexity of the
computation. We describe here the the main features of the
algorithm, which are those used in the proof of \eqref{4.5}, while
the existence of the algorithm itself will be proved in Subsection
\ref{subsec:4.4}, by exploiting a graphical representation of
contours.

We will restrict in the sequel to configurations $\und T$ such
that $\mathcal R (\und T)$ is  a singleton. As mentioned, the
basic property of the algorithm which associates a tree  to $\und
T$, is that each node  of the tree is representative of a subset
$\und T'$ of $\und T$ such that $\mathcal R (\und T')$ is a
singleton.

The root corresponds to the full $\und T$. Moreover the collection
of all the triangles associated to all the individuals of an
offspring is the same as the set of triangles associated to the
parent, so that a branching is nothing else than a partition of
the triangles present in the branching node.  In particular mass
is conserved in a branching.

White triangles are associated to contours consisting of a single
triangle, such triangle must be maximal, i.e.\ not contained in
any other triangle of $\und T$. Take notice, however, that the
converse may not be true, as it may happen that a maximal triangle
is one of the triangles associated to a black triangle or a
sphere.  If a maximal triangle $T$ corresponds to a white
triangle, then the spheres in the white triangle are associated to
the contours of the configuration $\und T'$ made of all the
triangles of $\und T$ which are contained in $T$.

The next properties we mention establish a quantitative relation
between the ordering of the offsprings in the tree and the
location of the corresponding triangles. If a black triangle
generates $n\ge 2$ h-triangles labelled consecutively,  call
$\Ga_i$, $1\le i \le n$, the triangle associated to the $i$-th
h-triangle (by itself each $\Ga_i$ is a contour, hence the
notation).  Then the triangles $\{T(\Ga_i)\}$ (recall that
$T(\Ga)$ is the minimal triangle which contains all the triangles
forming $\Ga$) are consecutive from left to right and
\begin{equation}
   \label{4.6}
\text{dist}\big(T(\Ga_i),T(\Ga_{i+1})\big) \le c
\min\big\{|\Ga_i|^3,|\Ga_{i+1}|^3\big\},\qquad i=1,..,n-1
 \end{equation}

% se cambio la P.1: > \to \geq
%\blue{ \begin{equation}
%   \label{4.6}
%\text{dist}\big(T(\Ga_i),T(\Ga_{i+1})\big) < c
%\min\big\{|\Ga_i|^3,|\Ga_{i+1}|^3\big\},\qquad i=1,..,n-1
% \end{equation}}
Moreover, if there are $k_i$ spheres between the $i$-th and
$(i+1)$-th h-triangle, call $\Ga^{(i)}_j$, $j=1,..,k_i$, the
contours associated to these spheres.  Then all $T(\Ga^{(i)}_j)$
are in between $T(\Ga_i)$ and $T(\Ga_{i+1})$, $\{T(\Ga^{(i)}_j)\}$
is sequential and the following constraint on their mutual
distances holds.  Letting $x_{\pm}(\Ga)$ as in the beginning of
Subsection \ref{subsec:4.1},
 the set of endpoints $\{x_{\pm}(\Ga^{(i)}_j), a
= x_+(\Ga_i), b= x_-(\Ga_{i+1})\}$ is such that  there is $p_i:
0\leq p_i\leq
k_i$ so that %with
% \begin{eqnarray}
%   \label{4.7}
%&&\text{dist}\Big(T(\Ga^{(i)}_j),\{T(\Ga^{(i)}_{j-1}),T(\Ga^{(i)}_{j+1})\Big)
%\le c |\Ga^{(i)}_j|^3,\qquad j=1,..,k_i\\&& \text{%where \hskip2cm
%$ T(\Ga^{(i)}_{0})\equiv
%T(\Ga_i);\;T(\Ga^{(i)}_{k_i+1})=T(\Ga_{i+1})$} \nonumber
% \end{eqnarray}

%\begin{eqnarray}
%   \label{4.7}
%&& 0 \le x_-(\Ga^{(i)}_1)-a\le  c |\Ga^{(i)}_1|^3, \;\;\;\;\; 0 \le
%x_-(\Ga^{(i)}_2)- x_+(\Ga^{(i)}_1)\le  c |\Ga^{(i)}_2|^3,\dots\nonumber\\
%&& \hskip 3cm \dots, x_-(\Ga^{(i)}_{p_i})- x_+(\Ga^{(i)}_{p_i-1})\le c
%|\Ga^{(i)}_{p_i}|^3
%\nonumber\\
%&& 0 \le b- x_+(\Ga^{(i)}_{k_i})\le  c |\Ga^{(i)}_{k_i}|^3, \;\;\;\;\; 0 \le
%x_+(\Ga^{(i)}_{k_i})- x_-(\Ga^{(i)}_{k_i-1})\le  c
%|\Ga^{(i)}_{k_i-1}|^3,\dots \nonumber\\
%&& \hskip 3cm \dots, x_+(\Ga^{(i)}_{p_i+1})- x_-(\Ga^{(i)}_{p_i+2})\le  c
%|\Ga^{(i)}_{p_i+1}|^3
% \end{eqnarray}
 \begin{eqnarray}
   \label{4.7}
&& 0 \le x_-(\Ga^{(i)}_1)-a\le  c |\Ga^{(i)}_1|^3{+1}, \;\;\;\;\; 0 \le
x_-(\Ga^{(i)}_2)- x_+(\Ga^{(i)}_1)\le  c |\Ga^{(i)}_2|^3{+1},\dots\nonumber\\
&& \hskip 3cm \dots, x_-(\Ga^{(i)}_{p_i})- x_+(\Ga^{(i)}_{p_i-1})\le c
|\Ga^{(i)}_{p_i}|^3{+1}
\nonumber\\
&& 0 \le b- x_+(\Ga^{(i)}_{k_i})\le  c |\Ga^{(i)}_{k_i}|^3{+1}, \;\;\;\;\; 0 \le
x_+(\Ga^{(i)}_{k_i})- x_-(\Ga^{(i)}_{k_i-1})\le  c
|\Ga^{(i)}_{k_i-1}|^3{+1},\dots \nonumber\\
&& \hskip 3cm \dots, x_+(\Ga^{(i)}_{p_i+1})- x_-(\Ga^{(i)}_{p_i+2})\le  c
|\Ga^{(i)}_{p_i+1}|^3{+1}
 \end{eqnarray}
Finally if $\Ga_j$, $j=1,..,k$, are the sets of triangles associated to the spheres
inside a white triangle, represented by $T$, then the triangles $T(\Ga_j)$ satisfy the
analogue of \eqref{4.7}
%%{with $a=x_-(T)-1$ and $b=x_+(T)+1$.} %se cambio la P.1
with $a=x_-(T)+1$ and $b=x_+(T)-1$. %se lascio le P.1
These are the only properties on the structure of contours that we
will use in the proof of \eqref{4.5} in Subsection
\ref{subsec:4.6}, next subsection is only an existence proof of
the algorithm for associating a tree to a contour with the
properties we have been describing so far, and, to a first
reading, it may be skipped.

\vskip2cm

 \subsection{Graphical construction}
 \label{subsec:4.4}

We will construct here an algorithm which associates a tree (with
the properties described in the previous subsection) to any $\und
T$ such that $\mathcal R(\und T)$ is a single contour. The
algorithm is obtained via a graphical representation of $\und T$,
where we draw at any integer time $t \in \{0,1,..\}$ a
configuration of mutually disjoint squares with a side in $\mathbb
R$, called the basis of the square. Each square $S$ is
representative of a cluster $\{\und T\}_S$ of triangles in $\und
T$, with the property that $\mathcal R(\{\und T\}_S)$ is  a
singleton; the name ``squares'' is just to avoid confusion with
the original triangles and the $h$-triangles of the tree. The mass
of a square $S$ equals the sum of the masses of the triangles in
$\{\und T\}_S$ (the mass of a triangle being the number of
integers contained in its basis). The configurations of squares at
the different times will be viewed as the successive applications
of a renormalization group transformation.

\vskip.5cm

 \centerline {{\it The time $t=0$ configuration}}
 \nopagebreak
This is obtained by associating to each ``maximal'' triangle of
$\und T$ a square with same basis:  $T\in \und T$ is maximal if it
is not contained in any other triangle of $\und T$. By definition
of maximality the set of maximal triangles, hence of squares, is
sequential. The cluster of triangles $\{\und T\}_S$ represented by
the square $S$ consists of a maximal triangle $T$ and of all the
triangles contained in $T$. The mass of $S$, according to the
general rule, is then the sum of all the masses in $\{\und T\}_S$.
Statement (ii) in Lemma \ref{lemma1e} below, proves that these
squares verify the property that the sets $\{\und T\}_S$ form a
single contour, thus our definition of the square configuration at
$t=0$ is well posed. In Appendix \ref{app:garbage} we will prove:

\vskip.5cm

   \begin {lemma}
   \label {lemma1e}
Let $S$ be a square corresponding to a maximal triangle $T$. Then:
(i),\; $\mathcal R(\{\und T\}_S\setminus T)=\{\Ga_j\}$ is
sequential and the sequence $\{T(\Ga_j)\}$ satisfies the analogue
of \eqref{4.7} with $a=x_-(T)$ and $b=x_+(T)$; moreover, (ii),\;
$\mathcal R(\{\und T\}_S)$ consists of a single contour.

   \end {lemma}

\vskip.2cm

By (i) the sequence $\{T(\Ga_j)\}$ satisfies the same properties
as the sequence of triangles obtained from the spheres attached to
a white triangle, as described in the previous subsection.
Together with (ii), this shows that each one of the squares  at
time $t=0$ is a candidate for being a white triangle. Whether this
will really happen, does depend in a complex way on the relative
positions of the other triangles of $\und T$, as we will see after
completing the construction of the square process.

\vskip.5cm

 \centerline {{\it The next time-step configuration}}
 \nopagebreak
The construction of the configuration of squares at time $t=n+1$
 only depends on the configuration at time $t=n$, namely on the
location of the squares in the configuration and on their masses. Like at time $t=0$,
each square $S$ is representative of a collection  $\{\und T\}_S$ of triangles in
$\und T$, more and more complex as time increases, but, as said, the construction of
the configurations at the successive time will only depend on locations and masses of
the squares, the latter being the sum of all the masses of the triangles represented.
The rule for constructing the configuration at time $t=n+1$ given the one at time
$t=n$, defines the action of the renormalization group transformation mentioned at the
beginning of the subsection.

We start by drawing oriented arrows between pairs of squares: we put an arrow $(S,S')$
from $S$ to $S'$  if $|S|\le |S'|$, $|S|$ the mass of the square $S$ (in case of
equality if $S$ is before $S'$, going from left to right) and if the distance between
$S$ and $S'$ is $\le c |S|^3$. Arrows define a connection, which will be referred to
as a-connection (a for arrow), to distinguish it from the connection used in the
definition of contours. Two squares are a-connected if they can be joined via a chain
of pairs of squares, each pair linked by an arrow (independently of the direction of
the arrow).

To each a-connected component we associate a proto-square, which
is the minimal square which contains all the squares in that
component, we call them proto-squares because some of the
proto-squares will become a square in the configuration at time
$t=n+1$.  We will prove below that any two such proto-squares are
either disjoint or one contained in the other. We call maximal
those which are not contained in any other one. {\it The maximal
proto-squares are   the squares at time $t=n+1$}. The set $\{\und
T\}_S$ represented by  a maximal proto-square $S$, is the
collection of all $\{\und T\}_{S'}$, with $S'$ running over all
the squares at time $t=n$ which are contained in $S$.  By
maximality the new squares at time $t=n+1$ are sequential.  In
Lemma \ref{lemmaC.2}, we will prove that $\mathcal R(\{\und
T\}_S)$ consists of a single contour, thus legitimating  the
present definition of the square configuration at time $t=n+1$.

We next state some features of the construction needed later for
the identification of a tree structure. To this end it is
convenient to erase some arrows, thus we will call ``old arrows''
the arrows defined so far and (old a)-connected squares connected
by old arrows.  Old arrows are erased with the following rule: if
there are several arrows emanating from a same square, all in a
same direction (i.e.\ right or left), we keep only the minimal one
and erase all the others. This is done for all squares and all
directions. The arrows which are left are the new arrows, and we
will call (new a)-connected, squares connected by the new arrows.
In Lemma \ref{lem:A1}, it is proved that a set is (new
a)-connected iff it is (old a)-connected.    We will hereafter in
this section call arrows the new arrows and a-connected, (new
a)-connected sets.

The {\it ``shadow"} of the arrow $(S,S')$ is the interval between
%(in the sense of Definition\ref{def:interval-connect})
the two endpoints of $S$ and $S'$ which face each other.  If two
shadows   have non empty intersection, they must be one contained
in the other,   Lemma \ref{prop:A2}; such a statement proves the
above property about the fact that the proto-squares are either
disjoint or one contained in the other.

We then call {\it primary} an arrow with maximal shadow, i.e.\ which is not contained
in any other shadow, and {\it primary} the two squares connected by a {\it primary}
arrow. The set $\{\und T'\}$ of all triangles in $\und T$ whose basis are contained in
the shadow $(a,b)$ of a primary arrow $(S_1,S_2)$ is such that:

\vskip.5cm

   \begin {lemma}
   \label {lemma1ae}
  With the above notation, \;\;(i),\; $\mathcal R({\und T'})=\{\Ga_j\}$
 and $\{T(\Ga_j)\}$  is a sequence which satisfies the analogue of
\eqref{4.7} with $a=x_+(S_1)$ and $b=x_-(S_2)$ supposing for
instance that $S_1$ is before $S_2$; moreover, (ii),\; $\mathcal
R(\und T'')$  consists of a single contour,  $\und T''$ being the
union of all triangles in $\und T'$ and those associated to $S_1$
and $S_2$.

   \end {lemma}

\vskip.52cm

 Lemma \ref{lemma1ae} is  proved in Appendix
\ref{app:garbage}. The primary squares may thus become
h-triangles, as they form a sequence which satisfy \eqref{4.6},
while  all the squares in a shadow, called {\it secondary}, are
eligible for being the spheres which lie between two h-triangles
in the tree.

We finally observe that after a finite number of iterations the
process stabilizes, the final configuration consisting of a single
square $S$, $\{\und T\}_S=\und T$, its mass therefore being the
sum of all $|T|$ over $\und T$; the basis of $S$ is $T(\Ga)$, the
triangle representative of the contour $\Ga= \mathcal R(\und T)$.
Any other final state would in fact contradict the assumption that
$\mathcal R(\und T)$ consists of a single contour.

\vskip.5cm

 \centerline {{\it The tree structure}}
 \nopagebreak
We have constructed so far, for any contour, a process, called the
square process, evolving at integer times, whose state space is a
square configuration, each square with its own mass.  The
evolution consists of a clustering mechanism, for which a cluster
of squares at time $t$ becomes a single square at time $t+1$.  We
have also distinguished in a forming cluster some squares which
are primary, the others being called secondary.  Our purpose now
is to identify a tree structure from $\und T$ via the realization
of the square process. To this end, let $t_f$ be the first time
when the final configuration, consisting of a single square, is
reached. This is identified to the root of the tree we are going
to construct. If $t_f=0$ the root is a white triangle, otherwise
it is black. In the former case, the configuration at time 0 has
only one square, $S$, which,  recalling the definition, means that
there is a unique maximal triangle, $T$, in $\und T$, and $S$ has
same basis as $T$. The spheres attached to the white triangle root
of the tree, are identified to the contours $\mathcal R( \und T
\setminus T)$, by Lemma \ref{lemma1e} such an identification
respects the requests of Subsection \ref{subsec:4.3}.  Notice that
the identification of the spheres attached to a white triangle
requires the knowledge of $\und T$ and cannot be read only from
the square process, which, as we will see, only identifies white
and black triangles and spheres between h-triangles, all with
their masses, but it does not give any information on the
structure of the spheres inside the white triangles, except for
their masses.

If $t_f>0$, the root is a black triangle and its offspring is the
configuration at time $t_f-1$, identifying primary squares with
h-triangles and secondary squares with spheres, consistently with
the properties of such objects, by  Lemma \ref{lemma1ae}. Let $S$
be one of the primary squares and $t_S<t_f-1$ the time in the
square process when there is a cluster of more than one square
which at time $t_S+1$ becomes $S$; if such a time does not exist,
then $S$ was present also at time 0, and it is identified to a
white triangle with same procedure as above. Otherwise $S$ is
identified to a black triangle, whose offspring is determined by
the configuration of squares at time $t_S$ which merge into $S$ at
time $t_S+1$, with same rules as those described for the branching
of the root. By iterating the procedure we complete the
identification of the tree.

%Examples of the forward and backward processes are drawn in the
%Figures \ref{fig:upwards} and \ref{fig:downwards}
%

%
%\begin{figure}[h]%h=here t=top b=bottom p=Page of floats
%\centering \resizebox{7cm}{!} {\rotatebox{-0}{\includegraphics{branch.ps}}}
%%\caption{Upwards}
%\caption{}\label{fig:upwards}
%\end{figure}
%
%
%\begin{figure}[h]%h=here t=top b=bottom p=Page of floats
%\centering \resizebox{7cm}{!} {\rotatebox{-0}{\includegraphics{branch-back.ps}}}
%%\caption{Reading downwards the above picture}
%\caption{} \label{fig:downwards}
%\end{figure}
\vskip2cm

 \subsection{Proof of Theorem \ref{thm4.2}}
 \label{subsec:4.6}
%Since the number of sites $x\in \mathbb Z$ contained in the basis
%of a triangle is equal to its length and the sum of the lengths of
%the bases of the triangles in a contour is the mass of the
%contour,
Since the number of translates of a contour with the property tha
$0\in \Ga$, is bounded by $|\Ga|$, the proof of Theorem
\ref{thm4.2} reduces to proving that for any $b>0$ large enough
and any $m>0$,
 \begin{equation}
   \label{4.9}
G_m:=  \sum_{\Ga: |\Ga|=m,  x_-(\Ga)=0} w_b(\Ga) \le  2 e^{-b
m^\al}
 \end{equation}
The proof is by induction on the mass $m$ of the contour, recall that the mass of a
contour is necessarily an integer. We thus suppose \eqref{4.9} proved whenever $m\le
M-1$ and want to prove it for $M$.  We have
 \begin{equation}
   \label{4.10}
G_M= G'_M+ G''_M
 \end{equation}
 \begin{equation}
   \label{4.11}
G'_M=  \sumtwo{|\Ga|=M,  x_-(\Ga)=0}{\text{root of $\Ga$ is
white}} \;w_b(\Ga),\qquad G''_M= \sumtwo{|\Ga|=M,
x_-(\Ga)=0}{\text{root of $\Ga$ is black}}\; w_b(\Ga)
 \end{equation}

      \vskip .5cm

We start by bounding $G'_M$.  We call $\ell=|T(\Ga)|$,
 $n$ the number of contours (spheres) attached to the white
triangle; $m_1,..,m_n$ their masses.  These variables are not
independent, as, for instance, we must have
$m_1+\dots+m_n+\ell=M$.  We organize the sum in \eqref{4.11} by
fixing  $\ell$, $n$, $m_1,..,m_n$, then summing over all the
contours compatible with such specifications and with \eqref{4.7}
and finally summing over the specifications  $\ell$, $n$,
$m_1,..,m_n$.

Let $\Ga_1,..,\Ga_n$ be $n$ contours whose masses are $m_1,..,m_n$
and all with $x_-(\Ga_i)=0$.  We call $X_n(\ell,\Ga_1,..,\Ga_n)$
the set of all $(x_1,..,x_n)$ such that the collection
$\{S_{x_i}(\Ga_i)\}$, $S_x$ denoting translation by $x$, fulfills
\eqref{4.7} with $a=0$, $b=\ell$ and $k_i=n$.  We then have
     \begin{eqnarray*}
G'_M&\leq &\sum_{\ell>0} e^{-b\ell^\al}\sum_{n\ge 0}
\sumtwo{m_1,\dots,m_n}{m_1+m_2\dots+m_n+\ell=M}
\sumtwo{\Ga_1..\Ga_n}{|\Ga_i|=m_i,\;x_-(\Ga_i)=0}
|X_n(\ell,\Ga_1,..,\Ga_n)|\prod_{i=1}^n w_b(\Ga_i)
      \end{eqnarray*}
Writing $[a\wedge b]:=\min\{a,b\}$, we have
     \begin{eqnarray*}
&& |X_n(\ell,\Ga_1,..,\Ga_n)|\le (n+1)  \prod_{i=1}^n
[cm_i^3\wedge \ell]
      \end{eqnarray*}
where $n+1$ counts the number of values that $p_i$ can take when
$k_i=n$ in \eqref{4.7}. Using the induction assumption we then get
     \begin{eqnarray*}
G'_M&\leq &\sum_{\ell>0}\sum_{n\ge 0} \sumtwo{m_1,\dots,m_n}{m_1+m_2\dots+m_n+\ell=M}
\;2^n e^{-b(\ell^\al+m_1^{\al}+\dots+m_n^{\al})} (n+1)\prod_{i=1}^n [cm_i^3\wedge
\ell]
      \end{eqnarray*}
To select the maximal among all masses, we  rewrite the above as
    \begin{eqnarray}
G'_M&\leq & \sum_{\ell>0}\sum_{n\ge 0 } 2^n(n+1)
\sumtwo{m_1,\dots,m_n}{m_1+m_2\dots+m_n+\ell=M}
e^{-b(\ell^\al+m_1^{\al}+\dots+m_n^{\al})} \prod_{i=1}^n [c m_i^3\wedge \ell]\cdot \nonumber\\
&& \hskip3cm \Big[\Ii_{\{\ell\geq m_i \; \forall i \}}+ \Ii_{\left\{{m_1\geq m_i \;
\forall i\neq 1 \atop m_1\geq \ell }\right\}}+\dots+\Ii_{\left\{ {m_n\geq m_i \;
\forall i\neq n \atop m_n\geq \ell} \right\}}\Big]
      \label{x}
    \end{eqnarray}
In the term where the maximum is  $m_i$, we bound $[c m_i^3\wedge
\ell] = \ell \le c\ell^3$ and get
 %by Lemma \ref{lem:3}, we have:
    \begin{eqnarray}
    \nonumber
G'_M &\leq& \sum_{n\ge 0}2^n
(n+1)^2\sumtwo{x_1+\dots+x_n+y=M}{0\le x_i\leq y} e^{-b y^\alpha}
\prod_{i=1}^n [e^{-b x_i^\al }(c x_i^3)]
\\
&\leq& \sum_{n\ge 0}
(n+1)^2(2c)^{n}\sumtwo{x_1+\dots+x_n+y=M}{0\le x_i\leq y}
e^{-by^\al+ \sum_i^n(b-a)x_i^\al} \prod_{i=1}^n [e^{-a x_i^\al }
x_i^3]
%\sumtwo{x_1+\dots+x_n+y=M}{x_i\leq y}\prod_{i=1}^n
%e^{-a(x_i^\al)}(2x_i^3)
\label{eq:4.11b}
  \end{eqnarray}
By  Lemma \ref{lem:4}, if $b/a$ is large enough,
    \begin{eqnarray}
&& \exp\{-by^\al+ \sum_i^n(b-a)x_i^\al\} \le \exp\{-b[y+ x_1+\cdots +x_n]^\al\}
\label{4.z}
  \end{eqnarray}
so that
     \begin{eqnarray*}
G'_M&\leq& e^{-bM^\al}\sum_{n \ge 0}(n+1)^2(2c)^{n} \Big( \sum_x e^{-a
x^\al}x^3\Big)^n
%\sumtwo{x_1+\dots+x_n+y=M}{x_i\leq
%y}\prod_{i=1}^n e^{-a(x_i^\al)}(2x_i^3)
     \end{eqnarray*}
Calling $\delta(a)$ the sum in the last bracket and noticing that $\delta(a)\to 0$ as
$a\to \infty$,  for $a$ large enough
    \begin{eqnarray*}
G'_M&\leq& e^{-bM^\al}\Big(1+\sum_{n\ge 1
}(n+1)^2(2c)^{n}[\delta(a)]^n\Big) \leq \frac{ 3 }{2} ~e^{-bM^\al}
\end{eqnarray*}

    \vskip 1.5cm

\textbf{Bound on $G''_M$}. We now call $n\ge 2$ the number of
[black and white] triangles generated by the root, $m_i$ their
masses. We fix all the contours with such specifications and sum
over the spheres between two consecutive triangles. Denote by
$k_i\ge 0$ the number of spheres between the $i$-th and $(i+1)$-th
triangles; $m_j^i$ their masses. The space interval where such
spheres can be located is determined by the position of the
triangles $i$ and $i+1$, by \eqref{4.6} its length is bounded by
$cm_{i,i+1}^3$, where $m_{i,i+1}:=[m_i\wedge m_{i+1}]$.  Then the
sum over the spheres, once their number and masses are fixed, is
bounded as in \eqref{x}. We can also sum over all possible
realizations of the $n$ black and white triangles, given their
number and masses using \eqref{4.6} and the induction assumption.
We then get (with an extra $2^n$ factor counting the number of
ways to color, either black or white, the $n$ triangles)
    \begin{eqnarray*}
G''_M&\leq&\sum_{n\ge 2} 2^n \sumtwo{m_1,\dots,m_n}{m_i>0}
\left\{\prod_{i=1}^{n-1} c m^3_{i,i+1}\right\}
\left\{\prod_{i=1}^n [2e^{-b m_i^\al}] \right\} \\&&
 \hskip1cm
\times \sum_{k_1\ge 0} \sumtwo{m_1^{1}\dots m_{k_1}^1}{m_j^1>0}
\left\{(k_1+1)\prod_{j=1}^{k_1} c[(m_j^1)^3\wedge
m_{1,2}^3]\right\}
\left\{\prod_{j=1}^{k_1}[2e^{-b(m_j^1)^\al}]\right\} \cdots
 \\&&
 \hskip1cm
\cdots \sum_{k_{n-1}\ge 0} \sumtwo{m_1^{n-1}\dots
m_{k_1}^{n-1}}{m_j^{n-1}>0}
\left\{(k_{n-1}+1)\prod_{j=1}^{k_{n-1}} c[(m_j^{n-1})^3\wedge
m_{n-1,n}^3]\right\}
\left\{\prod_{j=1}^{k_{n-1}}[2e^{-b(m_j^{k_{n-1}})^\al}]\right\}
\\&&
 \hskip1cm \times
\Ii_{\{\sum_i m_i+\sum_{k,l} m_k^l=M\}}
    \end{eqnarray*}
We fix $n,k_1,\dots,k_{n-1}$ and sum over all masses. As in
\eqref{eq:4.11b} we split the sum by fixing which one of the
masses is larger.  This will give a factor $(n+k_1+\dots+k_{n-1})$
equal to the number of masses which are present.  Except for the
largest mass we write the generic factor $e^{-b m^\alpha}=
e^{-(b-a)m^\alpha}e^{-am^\alpha}$.
 In order to
apply Lemma \ref{lem:4} and get the analogue of \eqref{4.z}, we
must check that there is not a term with the maximal mass to the
cube. If the maximal is one of the masses $m_i$, then it does not
appear because we have products of $m_{i,i+1}^3$ which
automatically select the smaller and avoid the larger.  If the
maximal mass is one of those relative to  spheres, say $m^i_j$, we
use the same trick as for $G'_M$ and bound $[(m_j^{i})^3\wedge
m_{i,i+1}^3] \le m_{i,i+1}^3$, so that the term $(m_j^{i})^3$ does
not appear. Notice that in this way there could be   factors
$m_{i,i+1}^6$. We then get
 \begin{eqnarray*}
&\leq & e^{-b M^\al}\sum_{n\ge 2} 2^n\sum_{ k_1\ge 0
,\dots,k_{n-1}\ge 0
}(n+k_1+\dots+k_{n-1})(k_1+1)\cdots(k_{n-1}+1)\\&&\hskip3cm \times
\left(\sum_{x\geq
1}e^{-a x^\al}(2c)x^6\right)^{(n-1+k_1+\dots+k_{n-1})}%^{N-1}
   \end{eqnarray*}
calling $\delta =\sum_{x\geq 1}e^{-a x^\al}(2c)x^6$:
\begin{eqnarray*}
 &\leq& e^{-b
M^\al}\sum_{n\ge 2}2^n\delta^{n-1}\sum_{k_1\geq
0}\delta^{k_1}(k_1+1)\dots \sum_{k_{n-1}\geq
0}\delta^{k_{n-1}}(k_{n-1}+1)(n+k_1+\dots+k_{n-1})
     \end{eqnarray*}
Since $(n+k_1+\dots+k_{n-1})=[(k_1+1)+\cdots+(k_{n-1}+1)+1] \le
2[(k_1+1)+\cdots+(k_{n-1}+1)] $,
\begin{eqnarray*}
&\leq & e^{-b M^\al}\sum_{n\ge 2}2^n\delta^{n-1} 2n \Big( \sum_{k\geq
0}\delta^{k}(k+1)^2\Big)^n
\\
&\leq & e^{-b M^\al}\sum_{n\ge 2}\delta^{n-1} 2^{2n+1}n \leq \frac{\dis{e^{-b
M^\al}}}{2}
     \end{eqnarray*}
because for $a $ large enough $\sum_{k\geq 0} \delta^k (k+1)^2\leq
2$ and $\sum_{n\geq 2} \delta^{n-1} 2^{2n+1}n\leq  1/2$.  We have
thus proved that
   $$
   G_M= G'_M+G''_M\leq (\frac 32 +\frac 12)\; e^{-bM^\al}
   $$
hence \eqref{4.9} and Theorem \ref{thm4.2} are proved.

\vskip 2cm

 \centerline{{\bf Acknowledgments}}
We are indebted to Vladas Sidoravicius,  Herbert Spohn and Milos
Zahradnik for many helpful comments.

PAF and MC acknowledge kind hospitality resp.\ at the
 Universities of Roma Tor Vergata and S\~ao Paulo.

PAF   acknowledges financial support from FAPESP, CNPq and PRONEX;
 MC, IM and EP   from MURST;  IM and EP from NATO
 Grant PST.CLG.976552 and GNFM.

\vskip3cm

\appendix

\setcounter{equation}{0}

%\centerline{\bf Appendices} \vskip .5cm \noindent
\section{}

%\section{Bound on $W(L)$}

\label{app:A}

In this appendix we will prove \eqref{2.7}  identifying the value
of the parameter $\zeta$.  We set
  \begin{equation}
  \label{A.1}
          \zeta_\al:= 1 - 2(2^\al-1) >0,\quad \text{ $0< \al
          < \al_+:=\frac{\ln 3}{\ln 2} -1$}
  \end{equation}
observing that $\al_+>1/2$.  We call $W_\al(L)$ the r.h.s.\ of
\eqref{2.6}, the subscript underlining  the dependence on $\al$.

\vskip.5cm

      \begin{lemma}
      \label{lem:primo}
Given  $\al\in [0, \al_+)$, for  $J(1)$  large enough
  \begin{equation}
  \label{A.2}
 W_\al(L)\geq    \begin{cases}  \zeta_\al L^\al  & \text{if  $ \al>0$}
 \\
   2 \ln L +8& \text{if $\al=0$}
  \end{cases}
  \end{equation}

   \end{lemma}

\begin{proof}
We first consider the case $\al>0$.  Using \eqref{1.2},
\eqref{2.6} reads
\begin{eqnarray*}
W_\al(L)&=& \sum_{x=1}^{L}\left(\sumtwo{y\in [L+1,2L]}{y\in
[-L+1,0]}\frac{1}{|x-y|^{2-\al}}-
\sumtwo{y\in [2L+1,\infty]}{y\in [-\infty,-L]}\frac{1}{|x-y|^{2-\al}}\right)+2 (J(1)-1)\\
&
=&2\sum_{x=1}^{L}\left(\sum_{y=L+1}^{2L}\frac{1}{|x-y|^{2-\al}}-\sum_{y=2L+1}^{\infty}
\frac{1}{|x-y|^{2-\al}}\right)+2 (J(1)-1)\\
& =&2\sum_{x=1}^{L}\left(\sum_{y=L+1-x}^{2L-x}\frac{1}{y^{2-\al}}-
\sum_{y=2L+1-x}^{\infty}\frac{1}{y^{2-\al}}\right)+2 (J(1)-1)\\
\end{eqnarray*}
%
%\begin{eqnarray*}
%W_\al(L)&=& \sum_{x=0}^{L}\left(\sumtwo{y\in [L+1,2L]}{y\in
%[-L,-1]}\frac{1}{|x-y|^{2-\al}}-
%\sumtwo{y\in [2L+1,\infty]}{y\in [-\infty,-L-1]}\frac{1}{|x-y|^{2-\al}}\right)+2 (J(1)-1)\\
%&
%=&2\sum_{x=0}^{L}\left(\sum_{y=L+1}^{2L}\frac{1}{|x-y|^{2-\al}}-\sum_{y=2L+1}^{\infty}
%\frac{1}{|x-y|^{2-\al}}\right)+2 (J(1)-1)\\
%& =&2\sum_{x=0}^{L}\left(\sum_{y=L+1-x}^{2L-x}\frac{1}{y^{2-\al}}-
%\sum_{y=2L+1-x}^{\infty}\frac{1}{y^{2-\al}}\right)+2 (J(1)-1)\\
%\end{eqnarray*}
%
%
and using monotonicity to replace sums by integrals:
  \begin{eqnarray*}
%&\geq&\zeta' \int_0^L dx \left[\int_{L+1-x}^{2L-x}\frac{dz}{z^{2-\al}}-
%\int_{2L+1-x}^\infty \frac{dz}{z^{2-\al}}\right]\\
%&\geq&\zeta' \int_0^L dx \frac{1}{\al-1}\left\{\left[(2L-x)^{\al-1}-(L+1-x)^{\al-1}\right]+(2L+1-x)^{\al-1}
%\right\}\\
%&\geq&\zeta' \frac{1}{(\al-1)}\frac{1}{(\al)}\left[-2(2L-x)^{\al}+(L+1-x)^{\al}\right]_{0}^{L}
%\\
%&\geq&\zeta' \frac{1}{(1-\al)}\frac{1}{(\al)}\left[2L^{\al}-2(2L)^{\al}+(L+1)^{\al}-1\right]
%\\
%&\geq&\zeta' L^{\al}\frac{1}{(1-\al)(\al)}\left[3-2(2)^{\al}\right]
%\\
%&\geq&\zeta L^{\al}
%\\
%\\
%\\
&\geq&2 \sum_{x=1}^L
\left[\int_{L+1-x}^{2L+1-x}\frac{dz}{z^{2-\al}}-
\int_{2L-x}^\infty \frac{dz}{z^{2-\al}}\right]+2 (J(1)-1)\\
&\geq&2 \sum_{x=1}^L
\frac{1}{\al-1}\left\{\left[(2L-x)^{\al-1}-(L+1-x)^{\al-1}\right]+(2L-x)^{\al-1}
\right\}+2 (J(1)-1)
\\
&\geq&2 \sum_{x=1}^L
\frac{1}{1-\al}\left\{\left[-2(2L-x)^{\al-1}+(L+1-x)^{\al-1}\right]
\right\}+2 (J(1)-1)
\\
&\geq&\frac{2}{1-\al}\left[ -2\sum_{y=0}^{L-1}  (L+y)^{\al-1}+
\sum_{y=0}^{L-1} (y+1)^{\al-1}\right]+2 (J(1)-1)
\\
&\geq&\frac{2}{1-\al}\left[ -2\int_{y=-1}^{L-1}  (L+y)^{\al-1}+
\int_{y=0}^{L} (y+1)^{\al-1}\right]+2 (J(1)-1)
\\
&\geq&\frac{2}{\al(1-\al)}\left[ -2[(2L-1)^{\al}- (L-1)^{\al}]+
(L+1)^\al-1\right]+2 (J(1)-1) \ge \zeta_\al L^{\alpha}
\end{eqnarray*}
The last inequality holds for $L$ large enough if  $\alpha \in
(0,\al_+)$, and for all $L$ if $J(1)$ is large enough.
%(We could
%actually take $\alpha$ up to $\ln 3/\ln 2-1$).
%
% for $\al>3/2$ ($\al<1/2$) $\zeta$ is positive
%(the limit value is actually $\al< \frac{\ln 3/2}{\ln 2}=\frac{\ln 3}{\ln 2}-1$).

\vskip .5cm \noindent In the case $\al=0$ we can repeat the same computations
obtaining:
\[
W_0(L)
%\geq 2\left[ -2[\ln(2+\frac{1}{L})]+ \ln(L+2)+
%(J(1)-1)\right] %\ge 2\ln{L}+4
%\\
\geq 2\ln(L+2)+\left[2J(1) -4\ln(3)-2\right] \ge 2\ln{L}+8
\]
for $J(1)$ large enough.%$J(1)>7,5$.
The lemma is proved.
\end{proof}

\vskip 2cm \noindent

\section{}

%\section{Proof of Theorem \ref{thm4.1}}
\label{app:thm4}
 We start by a preliminary lemma.

 \vskip1cm

 \begin{lemma}
 \label{lemma4.1}

Let $\mathcal R(\und T)$ satisfy P.0, P.1 and P.2, $\Ga \in
\mathcal R(\und T)$ and $\und T'$ the configuration of triangles
in $\Ga$. Then $\mathcal R(\und T')=\{\Ga\}$.

 \end{lemma}

 \vskip.5cm

{\bf Proof.}  Writing $\mathcal R(\und T)=\{\Ga_i,i=1,..,n\}$,
denote by $\und T^{(i)}$ the triangles in $\Ga_i$ and write
   $$
\mathcal R(\und T^{(i)}) =
\Big\{\Ga_1^{(i)},\dots\Ga_{n_i}^{(i)}\Big\}
   $$
Each pair $(\Ga_i,\Ga_j)$, $i\ne j$, verifies P.1,
%\eqref{4.1}--\eqref{4.2},
we want to show that P.1 is also  verified by each distinct pair
$\Ga_j^{(i)}, \Ga_{j'}^{(i')}$.  This is by definition if $i=i'$,
let us then suppose $i\ne i'$. If $T(\Ga_i) \sqcap
T(\Ga_{i'})=\emptyset$, then the same holds for $T(\Ga_j^{(i)})$
and $T( \Ga_{j'}^{(i')})$ and \eqref{4.1} holds. If instead
$T(\Ga_i) \sqsubset T(\Ga_{i'})$ (or viceversa)
% $T(\Ga_j^{(i)})
%\sqsubset T(\Ga_i) \sqsubset T'$ for any $T'$ in $\Ga_{i'}$ and
%henceforth for any $T'$ in $\Ga_{j'}^{(i')}$.  Moreover
%\eqref{4.2} holds for $\Ga_j^{(i)}, \Ga_{j'}^{(i')}$, so that P.1.
%is verified.

If instead $T(\Ga_i) \sqsubset T(\Ga_{i'})$ (or viceversa) then
$\dist (\Ga_j^{(i)},\Ga_{j'}^{(i')})\geq\dist
(\Ga_j^{(i)},\Ga_{i'})\geq \dist (\Ga_i,\Ga_{i'})\geq c|\Ga_i|^{3}
\geq c|\Ga_j^{(i)}|^{3}\geq
c\min\{|\Ga_j^{(i)}|^{3},|\Ga_{j'}^{(i')}|^{3}\} $  so that
$\Ga_j^{(i)}, \Ga_{j'}^{(i')}$ verifies P.1.

 By applying P.2, $\mathcal R(\und
T)=\{ \Ga_j^{(i)}\}$ which must therefore coincide with
$\{\Ga_i\}$. Hence the decomposition of each $\Ga_i$ into
$\Ga_j^{(i)}$ is trivial, i.e.\ $n_i=1$ and $\Ga_1^{(i)}=\Ga_i$.
The lemma is proved.  \qed

 \vskip1cm

 {\bf Proof of Theorem \ref{thm4.1}.}
Uniqueness.  Suppose there are two algorithms, $\mathcal R^{(i)},
i=1,2$ which both satisfy P.1 and P.2 and let $\mathcal
R^{(i)}(\und T)=\{\Ga_j^{(i)}, j=1,...,n_i\}$. Let
 \begin{equation}
    \label{4.3}
A^{(1)}_h = \Ga_1^{(1)} \sqcap \Ga_h^{(2)}
 \end{equation}
be the collection of those triangles which are both in
$\Ga_1^{(1)}$ and $\Ga_h^{(2)}$. Of course the union of
$A^{(1)}_h$ over $h$ is equal to $\Ga_1^{(1)}$.

Call $ \{K^{(1)}_{h,j}, j=1,..,m_h\}= \mathcal R^{(1)}(A^{(1)}_h)
$.
% be the set of contours relative to the configuration of triangles
%in $A^{(1)}_h$, obtained using the rule $\mathcal R^{(1)}$.
Each distinct pair $K^{(1)}_{h,j},K^{(1)}_{h',j'}$  verifies P.1,
by an argument similar to that used in the proof of Lemma
\ref{lemma4.1} and which is omitted. Then, by P.2, $
\{K^{(1)}_{h,j},h=1,..,n_2, j=1,..,m_h\}= \mathcal R^{(1)}(\und
T^{(1)})$, $\und T^{(1)}$  the collection of all triangles in
$\Ga_1^{(1)}$. By Lemma \ref{lemma4.1}  the decomposition is then
trivial, which means that $\Ga_1^{(1)}=\Ga_i^{(2)}$ for some $i$.
By iteration we then conclude that the two systems of contours
$\{\Ga_j^{(1)}\}$ and $\{\Ga_i^{(2)}\}$ are identical. Thus
$\mathcal R^{(1)}= \mathcal R^{(2)}$.

Existence.  Given a configuration $\und T$ of triangles, we call
$\mathcal C$ the collection of all partitions $\und
C=(C_1,..,C_n)$ of $\und T$ so that each pair $(C_i,C_j)$, $i\ne
j$,  verifies P.1 and P.0 (relative to $\und T$. $\mathcal C$ is
non empty as the trivial partition in a single atom  verifies P.0
and P.1 (as in that case there is nothing to check). We   order
$\mathcal C$ by setting $\und C\ge \und C'$ if the partition $\und
C$ is finer than $\und C'$.  We claim that $\mathcal C$ has a
unique maximal element, which will be called $\mathcal M(\und T)$;
we will then prove that $\mathcal M(\cdot)$ satisfies P.1 and P.2
and conclude the proof of existence.

The claim will follow from showing that the partition
 \begin{equation*}
%   \label{4.2a}
\und C \vee \und C' =\{ C_i\sqcap C'_j\},\qquad \und C=
(C_1,..,C_n),\;\und C'= (C'_1,..,C'_m)
 \end{equation*}
is in  $\mathcal C$, if also $\und C$ and $\und C'$ are in
$\mathcal C$.

Without loss of generality we must thus prove that any distinct
pair $ (C_i\sqcap C'_j, C_{i'}\sqcap C'_{j'})$ verifies the
alternatives in P.1.  By symmetry between the two clusters, we may
suppose $i\ne i'$.  If $T(C_i)\sqcap T(C_{i'})=\emptyset$, then
also $T(C_i\sqcap C'_j)\sqcap T(C_{i'}\sqcap C'_{j'})=\emptyset$
and \eqref{4.1} holds. Let us suppose (again without loss of
generality) that  $T(C_i)\sqsubset T(C_{i'})$, then for any $T_k
\in C_{i'}$, either $T(C_i)\sqsubset T_k$ or $T(C_i)\sqcap T_k
=\emptyset$. If $T_k \in C'_{j'}$, in correspondence with the
previous alternative, either $T(C_i\sqcap C'_j)\sqsubset T_k$ or
$T(C_i\sqcap C'_j)\sqcap T_k =\emptyset$. If instead $T_k \notin
C'_{j'}$, then $T_k\notin C_{i'}\sqcap C'_{j'}$ and there is
nothing to check. In conclusion the pair $ (C_i\sqcap C'_j,
C_{i'}\sqcap C'_{j'})$ verifies the alternatives (i)--(ii) and in
case the latter is verified,  \eqref{4.2} holds.

To complete the proof we must show that $\mathcal M(\und T)$
satisfies $P.2$. Let $\und T$ and $\und T^{(i)}:\sqcup \und
T^{(i)}=\und T$ be as in P.2 and suppose that the elements of
$\{\mathcal M(\und T^{(i)}), i=1,..,k\}$ satisfy
\eqref{4.1}--\eqref{4.2}. Suppose by
 contradiction that $\mathcal M(\und T)$ is not equal to $\{\mathcal M(\und
T^{(i)}), i=1..,k\}$, since the latter is in $\mathcal C$
(relative to $\und T$), $\mathcal M(\und T)$ must then be finer
than $\{\mathcal M(\und T^{(i)}), i=1..,k\}$. But then we would
have a finer partition of $\und T^{(i)}$, for some $i$, which
still verifies P.1.  We have thus reached a contradiction.

% Denote
%by $\und \Ga(\und T):=(\Ga_1,\dots,\Ga_n)$ the maximum element of $\mathcal C$ for the
%configuration of triangle $\und T$, and consider two configurations $\und T_1$ and
%$\und T_2$ such that
%for any pair $\Ga_i(\und T_1),\Ga_j(\und T_2)$ %,
%%$\Ga_i(\und T_1)\in \und\Ga(\undT_1)$, $\Ga_j(\und T_2)\in \und\Ga(\und T_2)$,
%\eqref{4.1}-\eqref{4.2} hold. Suppose that by contradiction $\und\Ga(\und
%T_1\sqcup\und T_2)\neq \{\und \Ga(\und T_1), \und \Ga(\und T_2)\}$, this cannot be
%because, in that case, the partition  $\tilde\Ga$ obtained by intersecting the two
%$\tilde\Ga:=\big[\und\Ga(\und T_1\sqcup\und T_2)\vee \{\und \Ga(\und T_1), \und
%\Ga(\und T_2)\}\big]$ is finer and then $\und\Ga(\und T_1\sqcup\und T_2)$ is not the
%maximum element.

The theorem is proved.
 \qed

  \vskip 2cm

   As a consequence of   P.1 and P.2 we have the following obvious property,
  namely that by adding triangles it cannot happen
   that contours split; the new triangles can either form separate contours or join
   other pre-existent ones and possibly cause them to merge.

 \vskip1cm

 \begin{lemma}{Monotonicity}
 \label{lemma4.2}

 Let $\und T$, $\und T'$ be two
configurations of triangles, $\und T\sqsubset \und T'$, then for
any $\Ga\in \mathcal R(\und T) $, there is $\Ga'\in \mathcal
R(\und T')$ so that $ \Ga\sqsubset \Ga'$.

 \end{lemma}

 \vskip.5cm

 {\bf Proof.}
 Let $\Ga_0\in \mathcal{R}(\und T)$,
$\mathcal{R}(\und T')=\{\Ga'_j\}$ and (recalling the notation in
\eqref{4.3}) $A_j:=\Ga_{0}\sqcap \Ga'_j$. We must prove that for
any $j$, either $A_j=\emptyset$ or $A_j=\Ga_0$. Suppose by
contradiction that this is not the case. We then consider the new
partition of $\und T$: $\und C:= \big[\mathcal{R}(\und T)
\vee\big\{ (\mathcal{R}(\und T)\setminus\Ga_0), A_1,\dots,
A_m\big\}\big]$. $\und C$ is then in $\cC$ (relative to $\uT$) and
is finer than $\mathcal{R}(\und T)$, which contradicts the fact
that $\mathcal{R}(\und T)$ is the unique, finest partition of
$\und T$ verifying P.0, P.1 and P.2. The lemma is proved. \qed

     \vskip 3cm

     \setcounter{equation}{0}
         \section{}

%      \section{Proof of Lemma \ref{lemma1e} and  \ref{lemma1ae}}
       \label{app:garbage}
 \nopagebreak
 An interval $[a,b]$ is compatible with $\und T$  if
$a$  is an endpoint of a triangle  of $\und T$, $b$ is also  an
endpoint of a triangle  of $\und T$ and for all $T\in \und T $,
$T\sqcap (a,b)$ is either void or equal to $(a,b)$.
%In the
%previous proof, $a$ and $b$ are the endpoints of a maximal
%triangle, while in the proof of (ii) of Lemma \ref{lemma1ae} below
%they are the endpoints of squares which are a-connected.

 \vskip1cm

  \begin{lemma}
  \label{lemmaC1}
Let $\und T'' \sqsubset \und T$, with $\mathcal R(\und T'')$ a
singleton;  $[a,b]$ a  $\und T''$-compatible interval;  $\und T'$
the collection of all triangles of $\und T$ with basis in $(a,b)$.
Then if $\mathcal R(\und T',\und T'')$ is not a singleton, also
$\mathcal R(\und T)$ is not a singleton.
  \end{lemma}

 \vskip.5cm

{\bf Proof.}
 Since $\cR(\uT'')$ is a singleton, by Lemma \ref{lemma4.2}
there is a contour $\Ga_0$ in $\cR(\uT',\uT'')$ which contains
$\uT''$. In order to prove the lemma we must consider the case
      $$
\mathcal R(\uT',\uT'') = \{\Ga_0,\dots \Ga_n\}, \qquad n\ge 1
      $$
Since $\Ga_i$, $i\ge 1$,  is distinct from $\Ga_0$, it is a subset
of $\uT'$ and therefore $T(\Ga_i)$ is strictly contained in
$(a,b)$.  Let
    $$
    \cR(
    \{\und T \setminus  \uT'\}\sqcup \Ga_0   ) =\{\Ga_1'\dots \Ga_k' \}
    $$
We claim that
 \begin{equation}
   \label{C.1}
\mathcal R(\und T)=\{\Ga_1,\dots,\Ga_n, \Ga_1'\dots \Ga_k' \}
 \end{equation}
hence $\mathcal R(\und T)$ is not a singleton and the lemma is
proved.

To prove \eqref{C.1}, it is enough to show that
$\{\Ga_1,\dots,\Ga_n, \Ga_1'\dots \Ga_k' \}$ satisfy properties
P.0 and P.1 in the definition of contours, see Section
\ref{sec:3}, because \eqref{C.1} would then follow from P.2.

P.0 is obviously satisfied.  Pairs $\Ga_i,\Ga_j$ and $\Ga'_i,\Ga'_j$   satisfy P.1 by
definition, it thus remain to check P.1 for pairs $\Ga_i,\Ga'_j$. By Lemma
\ref{lemma4.2}, $\Ga_0$ is contained in one of the contours $\Ga'_j$, say $\Ga'_1$,
and let us start from this case. Then dist$(\Ga_i,\Ga'_1)=$ dist$(\Ga_i,\Ga_0)$,
(because the triangles in $\Ga'_1\setminus \Ga_0$ are outside $[a,b]$, while $\Ga_0$
contains triangle(s) whose endpoints are $a,b$ and $\Ga_i$ has support inside
$(a,b)$).  Since the pair $\Ga_i,\Ga_0$ satisfies P.1 then also $\Ga_i,\Ga'_1$
satisfies P.1. For the same reason as before,  also for  $j>1$,
dist$(\Ga_i,\Ga'_j)\ge$ dist$(T(\Ga_i),\{a,b\})=$ dist$(\Ga_i,\Ga_0)> c |\Ga_i|^3$,
(the latter inequality by (ii) of P.1).  Hence  dist$(\Ga_i,\Ga'_j)> c \min \{
|\Ga_i|^3,|\Ga'_j|^3\}$. We have thus completed the proof of P.1: \eqref{C.1} and
Lemma \ref{lemmaC1} are thus proved. \qed

\vskip 1cm

{\bf Proof of (ii) of Lemma \ref{lemma1e}}.   We apply  Lemma
\ref{lemmaC1} with $\uT''=T$ (the maximal triangle), $a,b$ the
endpoints of $T$. Since  $\mathcal R(\und T'')=T$, $\mathcal
R(\und T'')$ is  a singleton; $\uT'= \{\uT\}_S\setminus T$. By
assumption $\mathcal R(\und T)$ is  a singleton, then, by Lemma
\ref{lemmaC1}, also $\mathcal R(\und T',\und T'')$ is a singleton,
hence (ii) of Lemma \ref{lemma1e} because  $\mathcal R(\und
T',\und T'')=\mathcal R(\{\und T\}_S)$.   \qed

 \vskip 1cm

{\bf Proof of (ii) of Lemma \ref{lemma1ae}}.  By definition of
squares, $\mathcal R(\{\und T\}_{S_i})$, $i=1,2$, are singletons;
then  $\mathcal R(\{\und T\}_{S_1}, \{\und T\}_{S_2})$ is a
singleton as well,
%, by monotonicity, Lemma \ref{lemma4.2}, and
because $S_1$ and $S_2$ are a-connected. We then apply Lemma
\ref{lemmaC1}, identifying $T'' =\{ \{\uT\}_{S_1},
\{\uT\}_{S_2}\}$ and $a,b$  as the  endpoints of the squares
$S_1,S_2$ which face each other.  The argument is hereafter the
same   as in the proof of (ii) of Lemma \ref{lemma1e}.  \qed

 \vskip1cm

  \begin{lemma}
  \label{lemmaC.2}
Let $ \und S $ be the square configuration at time $t=n$.  Call $
\und S' $ the collection of squares in a maximal a-connected
component of $ \und S $, $\{\und T\}_{\und S'})$ the set of
triangles represented by the squares in $\und S'$. Then $\mathcal
R( \{\und T\}_{\und S'})$ is a singleton and $ \und S'$ will be
represented by a square  in the square configuration at time
$t=n+1$.
  \end{lemma}

 \vskip.5cm

{\bf Proof.}  Suppose  $\und S'$ is not a singleton (otherwise the
statement of the lemma would trivially hold).  If $S_1\in \und S'$
there must be $S_2\in \und S'$ with $S_1$ and $S_2$ endpoints of
an arrow.  Then either $S_1$ is an endpoint of a primary arrow, or
it is in a shadow of a primary arrow.  By the assumed maximality
of  $ \und S' $ it then follows that  $ \und S' $ is made of a
sequence of squares each one connected by a primary arrow to the
successive one and all other squares contained in the shadow of
these primary arrows.

 By (ii)
of Lemma \ref{lemma1ae}, the collection of all the triangles in
the shadow of a primary arrow and those in the two squares
connected by the primary arrow form a single contour.  The
statement of the lemma then follows by monotonicity, Lemma
\ref{lemma4.2}.
 \vskip1cm

\vskip 1cm

 {\bf Proof of (i) of Lemma \ref{lemma1e}}.  We will first prove that
 $\Ga_1,\dots,\Ga_n$ are sequential, where
 \begin{equation}
   \label{C.2}
\{\Ga_1,\dots,\Ga_n \}=\mathcal R(\{\und T\}_S\setminus T)
 \end{equation}
Suppose by contradiction that $T(\Ga_i)\sqsubset T(\Ga_j)$, $i \ne
j$.  By property P.1 in the definition of contours, there is a
minimal contour $\Ga_k$ distinct from $\Ga_i$ such that
$T(\Ga_i)\sqsubset T(\Ga_k)$.  Let  $[a,b]$ be the smallest
interval containing $T(\Ga_i)$ and compatible with $\Ga_k$.  Let
$\und T'$ be the collection of triangles with basis in $(a,b)$.
Then $\mathcal R(\und T',\Ga_k)$ contains at least $\Ga_i$ and
$\Ga_k$ (by Lemma \ref{lemma4.2}), so that, by Lemma
\ref{lemmaC1},   $\mathcal R(\und T)$ is not a singleton, against
the assumption. Therefore   $\Ga_1,\dots,\Ga_n$ are sequential.

We will next prove the analogue of \eqref{4.7},  calling $a^*,b^*$
the endpoints of $T$,  shorthanding
$\{\Ga_i\}:=\{\Ga_1,\dots,\Ga_n \}$ and labelling the contours so
that   $\Ga_i$ is before $\Ga_j$ when $i<j$ (we have already
proved that  $\{\Ga_i\}$ is sequential). There is  $\Ga_j$ such
that $\dist( \Ga_j,\{a^*,b^*\})\equiv
\dist(T(\Ga_j),\{a^*,b^*\})\leq \ab|\Ga_j|^3$, otherwise all $
\Ga_i $ would be contours in $\mathcal R(\und T)$, by the argument
already used several times above.  Supposing for the sake of
definiteness that $\dist( \Ga_j ,a^*)\leq \ab|\Ga_j|^3$, we then
claim that
 \begin{equation}
   \label{C.3}
\dist( \Ga_1 ,a^*)\leq \ab|\Ga_1|^3
 \end{equation}
 Suppose by contradiction that
there is $1<k\le j$ so that  for any $i< k$, $\dist(\Ga_i,a^*)>
c|\Ga_i|^3$ while $\dist( \Ga_k ,a^*)\leq \ab|\Ga_k|^3$. This
would imply that, for any $i<k$, $\ab|\Ga_i|^3< \dist( \Ga_i ,a^*)
\leq \dist( \Ga_k ,a^*)\leq c|\Ga_k|^3$.  Thus $|\Ga_i|\leq
|\Ga_k| $ for any $i\leq k$. Since $\{\Ga_1,..,\Ga_n\}$ are
distinct contours, $\dist(\Ga_i,\Ga_k)>
c\min\{|\Ga_i|^3,|\Ga_k|^3\}=c|\Ga_i|^3$, for any $i<k$.

Call $\und T'$ the collection of all triangles in
$\{\Ga_1,..,\Ga_{k-1}\}$, $[a,b]=[a^*,x_-(\Ga_k)]$, $\und T''=
\{T,\Ga_k\}$. We can then apply Lemma \ref{lemmaC1} because
$\mathcal R(\und T'')$ is a singleton, since $\dist( \Ga_k
,a^*)\leq c|\Ga_k|^3$: then $\{\Ga_1,..,\Ga_{k-1}\}$ are contours
for $\mathcal R(\und T)$ which contradicts the assumption that the
latter is a singleton. \eqref{C.3} is proved.

%\vskip1cm
%
%We claim now that $\{\Ga_1,..,\Ga_{k-1}\}$ are contours for
%$\mathcal R(\{\und T\}_S)$, which would be the final contradiction
%because we have already proved (ii) of Lemma \ref{lemma1e}.  Let
%$\{\Ga'_1,..,\Ga'_\ell\}= \mathcal R(\{T, \Ga_k,..,\Ga_n\})$ and
%then proceed as in the proof {\tiny of (ii) of  Lemma
%\ref{lemma1e}. Details are omitted.} of Lemma $C_1$, identifying
%$\uT''= \Ga_k\sqcup T$, the interval $[a,b]$ of such Lemma with
%$[a, x_-(\Ga_k)]$, $\uT'\to \sqcup_{i=1}^{k-1} \Ga_i$, and
%recalling that each of them has a distance $\dist(\Ga_i,\uT'')\geq
%c|\Ga_i|^{3}$ $i=1,k-1$.
%   We have thus proved that $k$ must be equal to 1.

By \eqref{C.3},  $\cR(\{T\}_{\Ga_1},T)$ is  a singleton, so that
the previous analysis applies again with $T$ replaced by
$\{\Ga_1,T\}$ and $a^*$ replaced by $a^*_1=x_+(\Ga_1)$, showing
that if $\dist( \Ga_j ,a^*)\leq \ab|\Ga_j|^3$ with $j>2$, then
$\dist( \Ga_2 ,a^*_1)\leq \ab|\Ga_2|^3$.  By iterating the
argument we then conclude the proof of  (i) of Lemma
\ref{lemma1e}.  \qed

%the proof can be iterate to prove that $[x_+(\Ga_1),b]$, the
%argument applies to the contours $\{\Ga_2,\dots \Ga_n\}$,
%recalling that
%$\cR(\{T\}_{\Ga_1},T,\sqcup_{i=21}^{n}\{T\}_{\Ga_i})$ consist of a
%single contour, see remark above.

   \vskip1cm

 {\bf Proof of (i) of Lemma \ref{lemma1ae}}.  Since $\mathcal
R(\{S_1,S_2\})$ is  a singleton (see the   proof above of (ii) of
Lemma \ref{lemma1ae}) the previous applies unchanged with $a^*$
and $b^*$ the endpoints of $S_1$ and $S_2$ which face each other.
\qed

\vskip .5cm \noindent

\vskip 1.5cm \noindent

     \vskip 3cm

     \setcounter{equation}{0}

  \section{}
 \label{app:D}
  \nopagebreak

\begin{lemma}
\label{lem:A1} A squares configuration $\und S$ is (new
a)-connected iff it is (old a)-connected.
\end{lemma}

\vskip .5cm

{\bf Proof.}  If  $\und S$ is (new a)-connected, then it is also
(old a)-connected, as the new arrows are also old arrows. We thus
only need to prove the reverse implication: we suppose $\und S$
(old a)-connected but not  (new a)-connected and want to show that
this leads to a contradiction.

We call ``odd'' a pair $S$ and $S'$ of squares when there is an
old arrow between $S$ and $S'$ (denoted by $(S,S')_{\text{\rm
old}}$) while  $S$ and $S'$ are not (new a)-connected.  We will
first show that if $\und S$ is (old a)-connected but not (new
a)-connected then there exist odd pairs; we will then prove that
odd pairs ``can be shortened'' in the sense that if  $S$ and $S'$
is an odd pair, then there is another square $S''$ in between $S$
and $S'$ such that either $S$ and $S''$ or $S'$ and $S''$ is an
odd pair. The endless iteration of the argument leads to a
contradiction.

{\it Existence of odd pairs.}   If $\und S$ is not  (new
a)-connected, there are two squares $S$ and $S'$ which are not
(new a)-connected; since $S$ and $S'$ are (old a)-connected, there
is a sequence $\{S_{\ell_i}, i=1,..,j\}$ such that each pair
$S_{\ell_i},S_{\ell_{i+1}}$ is connected by an old arrow, and
$S_{\ell_1}=S$ and $S_{\ell_j}=S'$. Then one of the pairs
$S_{\ell_i},S_{\ell_{i+1}}$ must be odd, otherwise $S$ and $S'$
would be (new a)-connected.

{\it Shortening odd pairs.} Writing $S\prec S'$ if the square $S$
is before $S'$ (recall that a square configuration  is
sequential), we label $\und S$ so that $S_1\prec S_2\prec \dots
\prec S_n$.  Let $S_k,S_{m}$ be an odd pair and suppose, without
loss of generality, that $S_k\prec S_m$ and that the old arrow
which connects them goes from $S_k$ to $S_{m}$. The old arrow
which  connects  $S_k$ to $S_{m}$ is not a new arrow, otherwise
$S_k$ and $S_m$ would be (new a)-connected, therefore there exists
$S_\ell: S_k\prec S_\ell \prec S_m; ~|S_\ell|\geq |S_k|$ such that
there is a new arrow from $S_k$ to $S_\ell$. Consider separately
the two possible cases $1)$ $|S_\ell|\leq |S_m|$ and $2)$
$|S_\ell|> |S_m|$.

 $1)$. Since $c|S_\ell|^{3}\ge
c|S_k|^{3}\ge\dist(S_k,S_m)>\dist(S_m,S_\ell)$, there is an old
arrow connecting $S_\ell$ and $S_m$; on the other hand, by
definition, $S_\ell$ and $S_k$ are (new a)-connected, hence
$S_\ell$ and $S_m$ cannot be (new a)-connected, hence $S_\ell,S_m$
is an odd pair.

 $2)$.  As in $1)$,
$c|S_m|^{3}>\dist(S_m,S_\ell)$, which implies that there is an old
arrow from $S_m$ to $S_\ell$, as well as an old arrow from $S_k$
to $S_\ell$.  There are two subcases, $a)$ $S_m$ and $S_\ell$ are
also connected by a new arrow or else $b)$ they are not.  In
subcase $a)$, $S_\ell$ is (new a)-connected to $S_m$, hence it
cannot be (new a)-connected to $S_k$, thus $S_k,S_\ell$ is an odd
pair.  In subcase $b)$, there is $S_h:  S_\ell\prec S_h \prec S_m;
~|S_h|\geq |S_m|$,  such that there is a new arrow from $S_m$ to
$S_h$.  Again, as $S_h$ is (new a)-connected to $S_m$,  it is not
(new a)-connected to $S_k$.
 On the other hand $|S_h|\geq |S_m|\geq |S_k|$, hence there
is an old arrow from $S_k$ to $S_h$, thus $S_k,S_h$ is an odd
pair.

This concludes the analysis of case $2)$, and the proof of the
shortening property of odd pairs.  Thus the lemma is proved.  \qed

  \begin{prop}
  \label{prop:A2}

The shadows of two new arrows have either empty intersection or
else, one is contained in the other.

   \end{prop}
%
%\begin{figure}[h]%h=here t=top b=bottom p=Page of floats
%\centering \resizebox{7cm}{!}
%{\rotatebox{-0}{\includegraphics{es4k.ps}}} \caption{This picture
%of  primary arrows  is not allowed (primary arrows are drawn in
%bold black) } \label{fig:es4}
%\end{figure}

\begin{proof}%[Proof of Proposition \ref{prop:2}]
%As in the previous case, we need the proof only for mutually
%external [heavy] triangles.
 Suppose by contradiction that  there
are four squares $S_a\prec S_u\prec S_b \prec S_z$, with the
crossing arrows $\vxi_{ab}$ (denoting a new arrow from $a$ to $b$)
and $\vxi_{uz}$.
%(we are then considering  the case $|S_a|\leq |S_b|$,
%$|S_u|\leq |S_z|$).
By definition of new arrow, this implies that there is no
arrow $\vxi_{au}$ (there could be however an arrow  in the
opposite direction $\vxi_{ua}$) and  that
%$\vxi_{ab}$ implies
$\dist(S_a,S_b)\leq c|S_a|^{3}$. Recalling that $S_u\prec S_b$
the only compatible sizes with the new arrow $\vxi_{a,b}$ are :
%then
%$\dist(S_a,S_u)<c|S_a|^{3}$ and since there is no  arrow
%$\vxi_{au}$ then
$|S_u|<|S_a|<|S_b|$. On the other hand,
$\dist(S_u,S_b)<\dist(S_u,S_z)\leq c|S_u|^{3}$ then, being $|S_b|> |S_u|$ there should be an
arrow $\vxi_{u,b}: S_u\to S_b$ contradicting the fact $\vxi_{uz}$
is a new arrow (i.e that $S_z$ is the first square connected with
$S_u$).

Consider now the case in which the crossing arrows are
$\vxi_{b,a}$ and $\vxi_{u,z}$ (that implies that $|S_b|>|S_a|$ and
$|S_z|>|S_u|$). The existence of these arrows implies that there
are no the arrows $\vxi_{b,u}$ and $\vxi_{u,b}$, and , since
$\dist(S_b,S_u)<\dist(S_a,S_b)\leq c|S_b|^{3}$, this implies that
$|S_u|\leq c|S_b|$. We get a contradiction by observing that, since
$\dist(S_u,S_b)\leq \dist(S_u,S_z)\leq |S_u|^3$, there should be
an arrow $\vxi_{u,b}$ that is incompatible with $\vxi_{u,z}$.

The other possible crossing cases are reduced to those above  by reflection and the
proposition is proved.

\end{proof}

\vskip 2.5cm \noindent

\setcounter{equation}{0}

\section{}

  \label{app:E}

   \begin{lemma}
     \label{lem:4}
Let  $\al\in [0,1/2]$, $a$ and $b$  positive and $b/a$  large
enough. Then for any $n\ge 2$, any $x_1,..,x_{n-1},y$ such that $1
\le x_i \le y$,
      \begin{equation}
       \label{eq:4}
b~ \h_\al(y)+(b-a)~  \sum_{i=1}^{n-1}\h_\al(x)\; \geq \; b~
\h_\al\big(\sum_{i=1}^{n-1}x_i+ y \big)
%
%b y^\alpha + (b-a) \big( x_1^\al +\dots+x_{n-1}^\al\big)\geq b
%\big(y+x_1+\cdots+x_{n-1}\big)^\alpha
%
\end{equation}
where $\h_\al(L)$ is defined in \eqref{2.7}.
\end{lemma}
\vskip .5cm

{\bf Proof.}  We will prove \eqref{eq:4} by induction on $n\ge 2$
showing that
   \[
f_n(x_1,..,x_{n-1},y):=\frac{b}{b-a}h_\al(y)+  \sum_{i=1}^{n-1}
\h_\al(x)\; - \frac{b}{b-a}\h_\al\big(y+\sum_{i=1}^{n-1} x_i\big)
   \]
%\[
%b~ \h_\al(y)+(b-a)~  \sum_{i=1}^{n-1}\h_\al(x)\; - \; b~
%\h_\al\big(\sum_{i=1}^{n-1}x_i+ y \big)
%   \]
is non negative in the set $1\le x_i \le y$.

We start the induction by supposing that for $n>2$, for any $2\le
m\le n$, $f_m\geq 0$ and want to prove that
$f_{n+1}(x_1,..,x_n,y)\geq 0$.  Since $f_{n+1}$ is symmetric in
the first $n$ variables, we may suppose, without loss of
generality, that $x_i \le x_n \le y \le x_1+\cdots+x_{n-1}+y=:L$.
Then
        \begin{eqnarray*}
f_{n+1}(x_1,..,x_n,y)&=& f_n(x_1,..,x_{n-1},y)
+h_\al(x_{n})+\frac{b}{b-a}h_\al(L)-
\frac{b}{b-a}h_\al(L+x_{n})\\
&=& f_n(x_1,..,x_{n-1},y)+f_2(x_n,L) \geq 0
           \end{eqnarray*}
To complete the induction we need to prove that $f_2(x,y) \geq 0$.

\vskip .5cm \noindent

The case $\al>0$.  We have
%We start the induction by supposing that for
%$n>2$, for any $2\le m\le n$, $f_m\geq 0$ and want to prove that
%$f_{n+1}(x_1,..,x_n,y)\geq 0$.  Since $f_{n+1}$ is symmetric in
%the first $n$ variables, we may suppose, without loss of
%generality, that $x_i \le x_n \le y \le x_1+\cdots+x_{n-1}+y=:L$.
%Then
%        \begin{eqnarray*}
%f_{n+1}(x_1,..,x_n,y)&=& f_n(x_1,..,x_{n-1},y)
%+x_{n}^\al+\frac{b}{b-a}L^{\al}-
%\frac{b}{b-a}[L+x_{n}]^{\al}\\
%&=& f_n(x_1,..,x_{n-1},y)+f_2(x_n,L) \geq 0
%           \end{eqnarray*}
%To complete the induction we need to prove that
        \begin{eqnarray*}
f_2(x,y) = y^\al g (x/y),\quad g(x) := x^{\al} +\frac{b}{b-a}-
\frac{b}{b-a}(x+1)^{\al},\;\; 0\le x \le 1
          \end{eqnarray*}
If $b/a$ is large enough, $g'(x)>0$ and $g(x)\geq g(0)=0$ and the
induction is proved.  Thus \eqref{eq:4} is proved in the case
$\al>0$.

%To conclude the proof we need to show that $f_2\ge 0$.  We have,
%as before,
%     \begin{eqnarray*}
%f_2&=& \frac{b}{b-a}y^{\al} +  x_1^{\al}-
%\frac{b}{b-a}(y+x_1)^{\al}
%\\
%& =&y^{\al}\left(\frac{b}{b-a}+  x^{\al}- \frac{b}{b-a}
%(1+x)^{\al}\right),\qquad x=x_1/y
%\\
%& =&y^{\al} g(x)\geq 0
%      \end{eqnarray*}

\vskip .5cm \noindent
      The case $\al=0$:

Let
    $$
    p:=\frac{b}{b-a}, \;\;\text{ choose $a$ so that $1<p<2$}
    $$
We have
   \begin{eqnarray*}
f_2(x,y) &=&p(\ln y+4) +  (\ln x + 4)-
p(\ln [x+ y ]-4)\\
&=&-p\ln\big(1+ x/y \big)+  \ln x  + 4
\\
&\geq& -2p  + 4 +   \ln x \geq 0
   \end{eqnarray*}
because $x\ge 1$ and $p<2$.

\qed

%
%
%
%    \begin{eqnarray*}
%f_n&=&p\ln y+4p +  \sum_{i=1}^{n-1}(\ln x_i) + 4(n-1)-
%p\ln [\sum_{i=1}^{n-1}x_i+ y ]-4p\\
%&=&-p\ln\left(\frac{\sum_{i=1}^{n-1}x_i}{y}+1\right)+
%\sum_{i=1}^{n-1}(\ln x_i) + 4(n-1)
%\\
%&\geq& -p\ln( n -1+1)  + 4(n-1)+  \sum_{i=1}^{n-1}(\ln x_i)
%   \end{eqnarray*}
%where we have exploited the fact that $x_i\leq y$ for any $i$
%\[
%\geq -pn +4n-4+  \sum_{i=1}^{n-1}(\ln x_i)\geq
%\sum_{i=1}^{n-1}(\ln x_i)\geq 0
%\]
%because $(4-p)n-4>2(n-2)\geq 0$ and $x_i\geq 1$ \vskip .5cm
%\noindent \qed

\vskip 2cm

 \setcounter{equation}{0}

\section{}

\label{app:F}

In this appendix we sketch the proof of the analogue of
\eqref{4.5} in the case $\al=0$, namely that
 \begin{equation}
   \label{F.1}
  \sum_{\Ga: |\Ga|=m, 0\in  \Ga} w^0_b(\Ga) \le 2 m e^{-b (\ln m+4)}
 \end{equation}
where
  \[
w_b^{0}(\Ga):=\prod_{T\in\Ga}e^{-b(\ln(|T|+4)}
=\prod_{T\in\Ga}\left(|T|^{-b}e^{-4b}\right)
  \]
\eqref{F.1} yields the analogue of \eqref{3.12}, i.e.\
\begin{equation}
   \label{F.2}
\mu_{\La}^+\Big( \{0\in  \Ga\}\Big) \le 2\sum _{m\ge 1} m e^{-
\beta (\ln(m)+4)}= 2~e^{-4\beta}\sum _{m\ge 1} m^{1-\beta}
 \end{equation}
The sum in \eqref{F.1} is bounded using  the same iterative
procedure as when $\al\in (0,1/2]$, with the fundamental
inequality \eqref{4.z} replaced  by the  ``convexity" inequality
  \[
b~ \h_0(y)+(b-a)~  \sum_{i=1}^{n-1}\h_0(x)\; \geq \; b~
\h_0\big(\sum_{i=1}^{n-1}x_i+ y \big)
  \]
proved in Appendix \ref{app:D}  for $0<a<b/2$. The proof of
\eqref{F.1} then follows closely that of \eqref{4.5} for $\al>0$,
and it is omitted.

  \qed

\vskip 3cm \noindent

\bibliographystyle{amsalpha}

\vskip 1.5cm \noindent
\end{document}